\documentclass[aps,prb,twocolumn,groupedaddress]{revtex4}
\usepackage[pointedenum]{paralist}
\usepackage[fleqn]{amsmath}
\usepackage{amssymb}
\usepackage{array, multirow}
\usepackage{dcolumn}
\newcolumntype{.}{D{.}{.}{10}}
\newcolumntype{,}{D{.}{.}{2}}
\usepackage{graphicx}
\usepackage{rotating}
\usepackage{pdfcolmk}
\usepackage[authormarkup=none,ulem=normalem,final]{changes}
\usepackage{xcolor}
\definecolor{mydarkblue}{rgb}{0.1,0,0.55}
\definecolor{darkgreen}{rgb}{0.1,0.5,0.1}
\definecolor{orange}{rgb}{1,0.4,0.0}
\usepackage{lmodern} 
\usepackage{textcomp}
\usepackage{comment}
\DeclareMathOperator \argmax {argmax}

\begin{document}

\bibliographystyle{apsrev4}

\title{
        Parametrization of  pair correlation function and static structure factor\\      of the one component plasma across coupling regimes
}
\author{N. Desbiens, P. Arnault, J. Cl\'erouin}
\affiliation{CEA, DAM, DIF, F-91297 Arpajon, France.}

\date{\today}
\begin{abstract}

We present a parametrization of the pair correlation function and the static structure factor of the Coulomb one component plasma (OCP) from the weakly coupled regime to the strongly coupled regime. Recent experiments strongly suggest that the OCP model can play the role of a reference system for warm dense matter. It can provide the ionic static structure factor that is necessary to interpret the x-ray Thomson scattering measurements, for instance. We illustrate this with the interpretation of a x-ray diffraction spectrum recently measured, using a Bayesian method that requires many evaluations of the static structure factor to automatically calibrate the parameters. For strongly coupled dusty plasmas, the proposed parametrization of the Coulomb OCP pair correlation function can be related to the Yukawa one, including screening. Further prospects to parametrize the static structure of Yukawa systems are also discussed.

\end{abstract}
\maketitle

A realistic description of hot dense plasmas requires a model that includes all many-body interactions explicitly in contrast to weakly coupled plasmas that are well represented by only binary collisions and mean-field effects. This regime is characterized by strong interactions between ions, outweighing their thermal kinetic energy and leading to a liquid-like structure. Such strongly correlated plasmas are found in a variety of environments including planetary interiors, \cite{BaraffeEtAl2010}  dwarf stars, \cite{KoesterEtAl1990} and neutron star crusts \cite{DaligaultEtAl2009} in astrophysics, and in many experimental set-ups of dusty plasmas, colloidal suspensions, \cite{BonitzEtAl2010} and warm dense matter (WDM) \cite{GrazianiEtAl2014} in inertial confinement fusion studies. \cite{LindlEtAl2004}  Liquid metals \cite{Shimoji1977} are also another manifestation of high Coulomb coupling in nature.

In strongly coupled plasmas, the many-body interactions are not amenable to a theoretical approach in perturbation since there is no small parameters available. Besides very demanding state-of-the-art molecular dynamics  simulations, a possible modeling comes through the definition of simpler idealized systems. These reference systems can be studied extensively by molecular dynamics once and for all, with results easily available through parametrizations and/or tabulations. The Coulomb \cite{Hansen1973,BausEtAl1980} and Yukawa \cite{HamaguchiEtAl1997} one-component-plasma (OCP) systems are good candidates of such reference models.

The Coulomb OCP represents a system of interacting ions, with the pair potential $V_C(r) = Q e/r$, in a neutralizing uniform background of electrons. All its static and dynamic properties depend on only one parameter, the Coulomb coupling parameter $\Gamma = Q^2 e^2 / a k_B T$, where $Q$ is the ionic charge, $T$ is the temperature, and $a = (4 \pi n/3)^{-1/3}$ is the Wigner-Seitz radius ($n$ is the ionic density). The Coulomb coupling parameter $\Gamma$  is a measure of the importance of correlation in plasmas as the ratio of the mean nearest neighbor interaction to the mean kinetic energy. The ideal gas-like behavior corresponds to $\Gamma\ll1$, liquid-like short-range order appears when $\Gamma\ge1$, and crystalline long-range order when $\Gamma\ge180$. Analytic parametrizations in $\Gamma$ are available for the equation of state \cite{Caillol1999,CaillolEtAl2010} and the ionic transport coefficients, viscosity \cite{Bastea2005,DaligaultEtAl2014} and diffusion. \cite{Daligault2006,Daligault2009} Interestingly, when the plasma is partially ionized, the coupling parameter $\Gamma$ can stay constant along isochores as a result of the increase of ionization compensating the increase of temperature. \cite{ClerouinEtAl2013,ArnaultEtAl2013,Clerouin2015,ClerouinEtAl2015a} In plasma mixtures, the coupling parameter $\Gamma$ may be different for each species giving rise to interesting coexistences between different coupling regimes. \cite{Arnault2013,WhitleyEtAl2015,TicknorEtAl2016}

In the Yukawa model, the polarization of the electrons close to each ion is accounted for by a screened potential $V_Y(r) = Q e \exp(-\kappa r)/r$, where $\kappa$ is an inverse screening length. This form of the pair potential originates from a linear response treatment of the electron gas in presence of a test charge, in the small wave number (long distance) limit, \cite{StantonEtAl2015} for given values of ionization, temperature, and density.

Here, we address the static structure of dense plasmas, as revealed by the pair distribution function (PDF) $g(r)$ and the static structure factor (SSF) $S(k)$. \cite{HansenEtAl2006} The PDF gives the proportion of ions around a given ion  as a function of their distance, with respect to the uniform distribution of a non-interacting system. The effect of repulsive interactions on the PDF is to create a correlation void at small distance, and corresponding peaks at larger distances reflecting the emerging shell structure, that ultimately forms the crystal lattice at solidification. The PDF is an input data in different theoretical frameworks: the variational theory of fluid \cite{BarkerEtAl1967,WeeksEtAl1971} or the thermodynamic integration for the equation of state; the quasi-localized charge approximation \cite{GoldenEtAl2000,GoldenEtAl2001} for the study of wave dispersions \cite{RosenbergEtAl1997} in strongly coupled plasmas. This structural information is also directly observable in dusty plasma experiments, \cite{BonitzEtAl2010} whereas the SSF $S(k)$, representing the Fourier content of $g(r)$, is probed by neutron and x-ray diffraction in WDM.

Despite many theoretical efforts to find analytical approximations of the OCP static structure from the integral equations of fluid theory, \cite{BausEtAl1980,ChaturvediEtAl1981,HansenEtAl2006} accurate enough SSF are only available as tabulations produced in the 80s. \cite{RogersEtAl1983} An alternative approach consists in using brute force to parametrize results of microscopic simulations.  Recently, Ott \textit{et al.} \cite{OttEtAl2014b,OttEtAl2015} performed large scale molecular dynamics simulations to extract analytic fits for the height of the first peak $g_\mathrm{max}$ of the Coulomb and Yukawa PDFs $g(r)$ and the radius of the correlation hole $r_{1/2}$, defined as the nearest distance to an ion where $g(r)=1/2$, reflecting the mutual repulsion of ions at short distance. 

In this paper, we present a parametrization of the PDF and the SSF of the Coulomb OCP as functions of the Coulomb coupling parameter $\Gamma$, based on extensive molecular dynamics simulations covering the whole fluid phase across coupling regimes ($0 \le \Gamma \le 180$). The PDF of the Yukawa system depends on the screening parameter $\kappa$ in addition to the coupling parameter $\Gamma$. Nevertheless, a correspondence between the Coulomb and Yukawa PDFs was evidenced recently. \cite{OttEtAl2014b} It defines an effective coupling parameter $\Gamma_\mathrm{eff}$, depending on $\Gamma$ and $\kappa$, that produces almost the same PDF. Indeed, we shall see that our parametrization of the OCP PDF can be used as well for the Yukawa systems.
 
The paper is organized as follows. The molecular dynamics simulations are first described in Sec.\,\ref{MD}. Then, the parametrizations of PDF (Sec.\,\ref{PDF}) and SSF (Sec.\,\ref{SSF}) are presented and their accuracy is assessed by direct comparison with simulations. Further constraints are also examined with the calculations of the equation of state using either the PDF or the SSF (Sec.\,\ref{EOS}). The relationship between Yukawa and OCP static structures is illustrated in Sec.\,\ref{YOCP} by the comparison of PDFs and SSFs corresponding to the same effective coupling parameter $\Gamma_\mathrm{eff}$, with emphasize on the limitations of the present parametrization \cite{OttEtAl2014b} of $\Gamma_\mathrm{eff}$ in the weak to moderate coupling regime. Finally, an example is given in Sec.\,\ref{XRTS} of the usefulness of our parametrization in the interpretation of x-ray Thomson scattering measurement. In this example, we use a Bayesian method to quantify the accuracy of this OCP-based representation of the ion feature.  Lastly, a summary and some prospects to extend the parametrization to the Yukawa system are presented in the conclusion (Sec.\,\ref{concl}).

\section{Molecular Dynamics Simulations}
\label{MD}

We used the ESPRESSO code \cite{Limbach2006704} to perform molecular dynamics simulations of the Coulomb OCP. This code allows one to use reduced units. In our simulations, the space variable $r$ is scaled by the Wigner-Seitz radius $a$, leading to a constant density of $n = 3/4 \pi$. The time unit is the inverse plasma frequency $\omega_P^{-1} = [4 \pi n Q^2 e^2/m]^{-1/2}$, where $m$ is chosen as the unit of mass for the ions. When the energy unit is chosen as $k_B T$, the pair potential energy reduces then simply to:
\begin{equation}
U_C(r) = \dfrac{\Gamma}{r}.
\end{equation}
We placed $N$ ions in a cubic box, that is periodically duplicated in each direction. Their dynamics is followed solving the Newton equations by a velocity Verlet algorithm, \cite{FrenkelEtAl2002} with (NVT ensemble) or without (NVE ensemble) a Berendsen thermostat. \cite{BerendsenEtAl1984} The long-range Coulomb forces between the ions, including their periodic images, are computed
using the particle-particle-particle-mesh method \cite{HockneyEtAl1988} that scales as $N\log N$. This method, also known as the P3M method, is efficiently parallelized in the ESPRESSO code, using one-dimensional fast Fourier transforms. 

After an equilibration of the system in NVT ensemble during 100\,$\omega_P^{-1}$, the PDF and the SSF are collected and averaged in NVE ensemble over a duration of 1000\,$\omega_P^{-1}$. Without thermostat (NVE), we also collect the velocity autocorrelation functions for a future study. As the temperature fluctuates in NVE, the targeted coupling is realized with a relative accuracy of $\Delta \Gamma / \Gamma = 1\%$. The Table \ref{tab:sim} gives for the different values of the coupling parameter $\Gamma$, the number $N$ of ions and the time step $\tau$ used.  Finite size effects have been checked by varying the timestep and the number of particles.

\begin{table}[!t]
\centering
\begin{tabular}{,,,>{~~~~~}c>{~~~}c}
\hline 
\multicolumn{3}{c}{$\Gamma$} &	$N$ 	&	$\tau$	\\
\multicolumn{1}{c}{from} & \multicolumn{1}{c}{to} & \multicolumn{1}{c}{step} & &  ($10^{-3}\omega_P^{-1}$) \\
\hline  
0.04 	& 0.09 	& 0.01	& 15000  &  1   \\
0.1 	& 0.9 	& 0.1	& 15000  &  1   \\
1 	& 4 	& 0.15	& 2500  &  5   \\
5 	& 155 	& 5	& 2500  &  5   \\
\hline 
\end{tabular}
\caption{Ranges of Coulomb coupling parameter $\Gamma$ studied in this work together with the number of particles $N$ and the timestep $\tau$ used in the MD simulations ($\omega_P$ is the plasma frequency).}
\label{tab:sim}
\end{table}

\section{Pair Distribution Function}
\label{PDF}

We used different functional forms to parametrize the PDF $g(r)$ at weak and strong couplings. For Coulomb coupling parameter $\Gamma$ less than 5, the weak coupling form $g_\mathrm{weak}(r)$ contains a non-linear Debye-H\"uckel contribution, \cite{BausEtAl1980} whereas at higher coupling, the strong coupling form $g_\mathrm{strong}(r)$ accounts for the oscillations after the first peak of $g(r)$ and their attenuations. The latter form is inspired from the works of Matteoli, Mansoori, \cite{MatteoliEtAl1995} and Lai \textit{et al.}:\cite{LaiEtAl2012}

\begin{figure}[!t]
\begin{center}
\includegraphics[width=7.7cm]{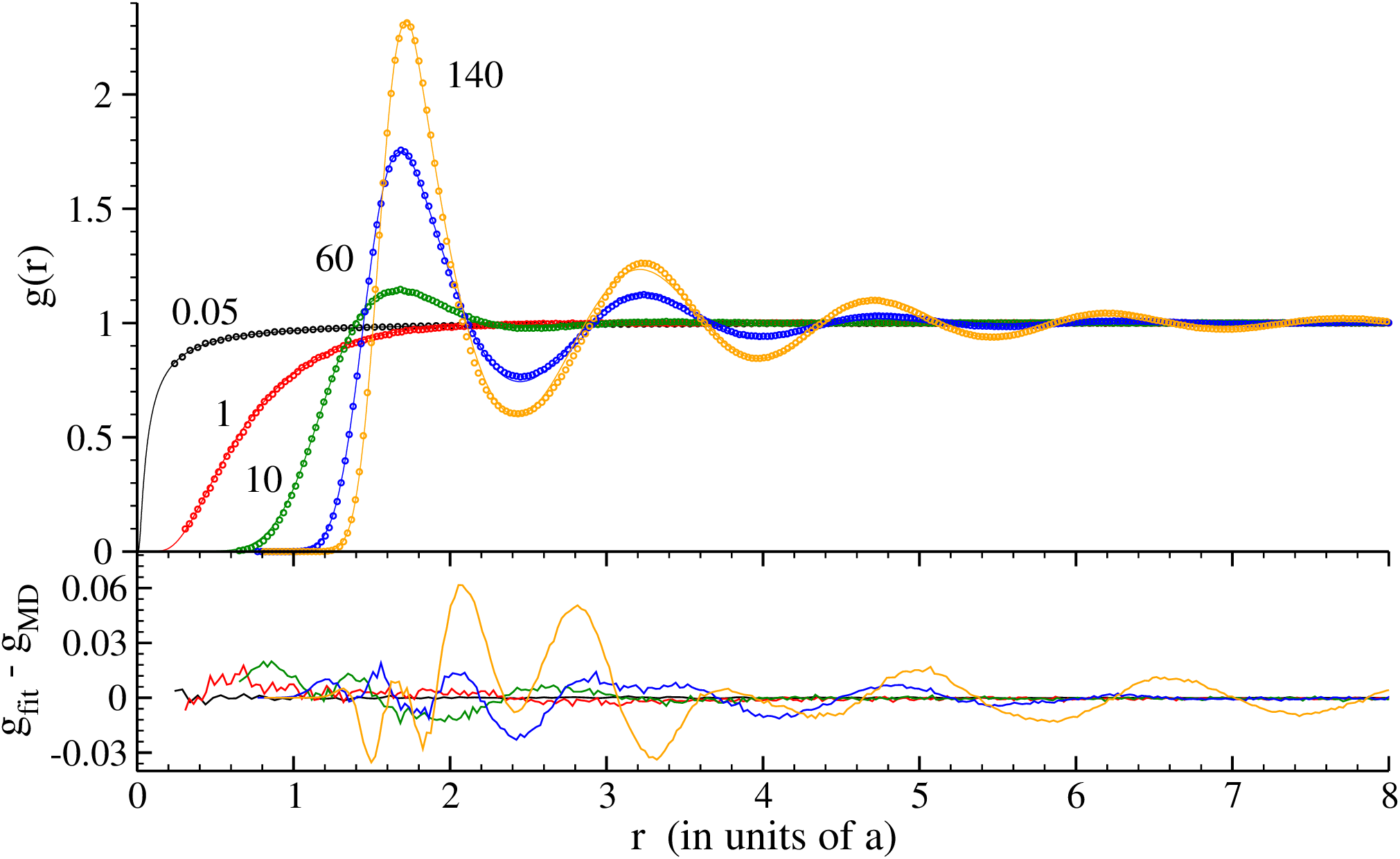}
\caption{Pair distribution functions for $\Gamma$ = 0.05 (black), 1 (red), 10 (green), 60 (blue) and 140 (orange). Top view: circles correspond to simulations while lines correspond to the parametrization. Bottom view: differences between fits ($g_\mathrm{fit}$) and simulations ($g_\mathrm{MD}$).} 
\label{fig:pdf}
\end{center}
\end{figure}

\begin{subequations}
\label{eq:pdf}
\begin{equation}
\label{eq:pdf1}
g_\mathrm{fit}(r) = 
\begin{cases}
g_\mathrm{weak}(r)	& \text{for}~\Gamma \le 5, \\
g_\mathrm{strong}(r)	& \text{otherwise},
\end{cases}
\end{equation}
\begin{equation}
\label{eq:pdf2}
g_\mathrm{weak}(r) = e^{-\frac{\phi}{r}\cdot e^{-\eta r}} + \chi\cdot e^{-\left(\frac{r-\rho}{\tau}\right)^2},
\end{equation}
\begin{equation}
\label{eq:pdf3}
g_\mathrm{strong}(r) = 
\begin{cases}
\sigma\cdot e^{-\mu (-x)^\nu}	& \text{for}~x\le0, \\
1 + (\sigma-1)\cdot\frac{\cos{\left(\alpha x + \beta \sqrt{x}\right)}}{\cosh{\left(x\cdot\Lambda(x)\right)}}	& \text{otherwise},
\end{cases}
\end{equation}
\begin{equation}
\label{eq:pdf4}
x = \frac{r}{\zeta} - 1,
\end{equation}
\begin{equation}
\label{eq:pdf5}
\Lambda(x) = (\delta - \epsilon)\exp{\left(-\sqrt{x/\gamma}\right)} + \epsilon.
\end{equation}
\end{subequations}

In the latter equations, and in the following, the distance $r$ is in units of the Wigner-Seitz radius $a$. A least-square fit method has been used to calibrate the parameters. At weak coupling, one gets:

\begin{eqnarray}
\phi & = & \Gamma \,(1 + 0.167\;\Gamma^{{4}/{3}}), \nonumber \\
\eta & = & \sqrt{3 \Gamma}\,(1 - 0.1495\;\sqrt{\Gamma}), \nonumber \\
\chi & = & 1.826\:10^{-2}\;\Gamma^{{4}/{3}}, \\
\rho & = & \min{\left\lbrace 2.50 \;;\; 1.378 + \frac{0.284}{\Gamma}\right\rbrace}, \nonumber \\
\tau & = & \min{\left\lbrace 1.25 \;;\; 0.505 + \frac{0.289}{\Gamma^{2/3}} \right\rbrace}. \nonumber
\end{eqnarray}

Parameters of strong coupling are:
\begin{eqnarray}
\zeta    & = & 1.634 + 7.934\:10^{-3}\;\sqrt{\Gamma} + \left(\frac{1.608}{\Gamma}\right)^2, \nonumber \\
\sigma   & = & 1 + 1.093\:10^{-2}\; {(\ln{\Gamma})}^3, \nonumber \\
\mu      & = & 0.246 + 3.145\;\Gamma^{{3}/{4}}, \nonumber \\
\nu      & = & 2.084 + \frac{1.706}{\ln\Gamma}, \nonumber \\
\alpha   & = & 6.908 + \left(\frac{0.860}{\Gamma}\right)^{{1}/{3}}, \\
\beta    & = & 0.231 - 1.785\;e^{-{\Gamma}/{60.2}}, \nonumber \\
\gamma   & = & 0.140 + 0.215\;e^{-{\Gamma}/{14.6}}, \nonumber \\
\delta   & = & 3.733 + 2.774\;\Gamma^{1/3}, \nonumber \\
\epsilon & = & 0.993 + \left(\frac{33.0}{\Gamma}\right)^{2/3}. \nonumber
\end{eqnarray}
It can be noticed that $\zeta$ and $\sigma$ are respectively the position and the height $g_\mathrm{max}$ of the first peak of $g(r)$, when $\Gamma > 5$.

\begin{figure}[!t]
\begin{center}
\includegraphics[width=7.7cm]{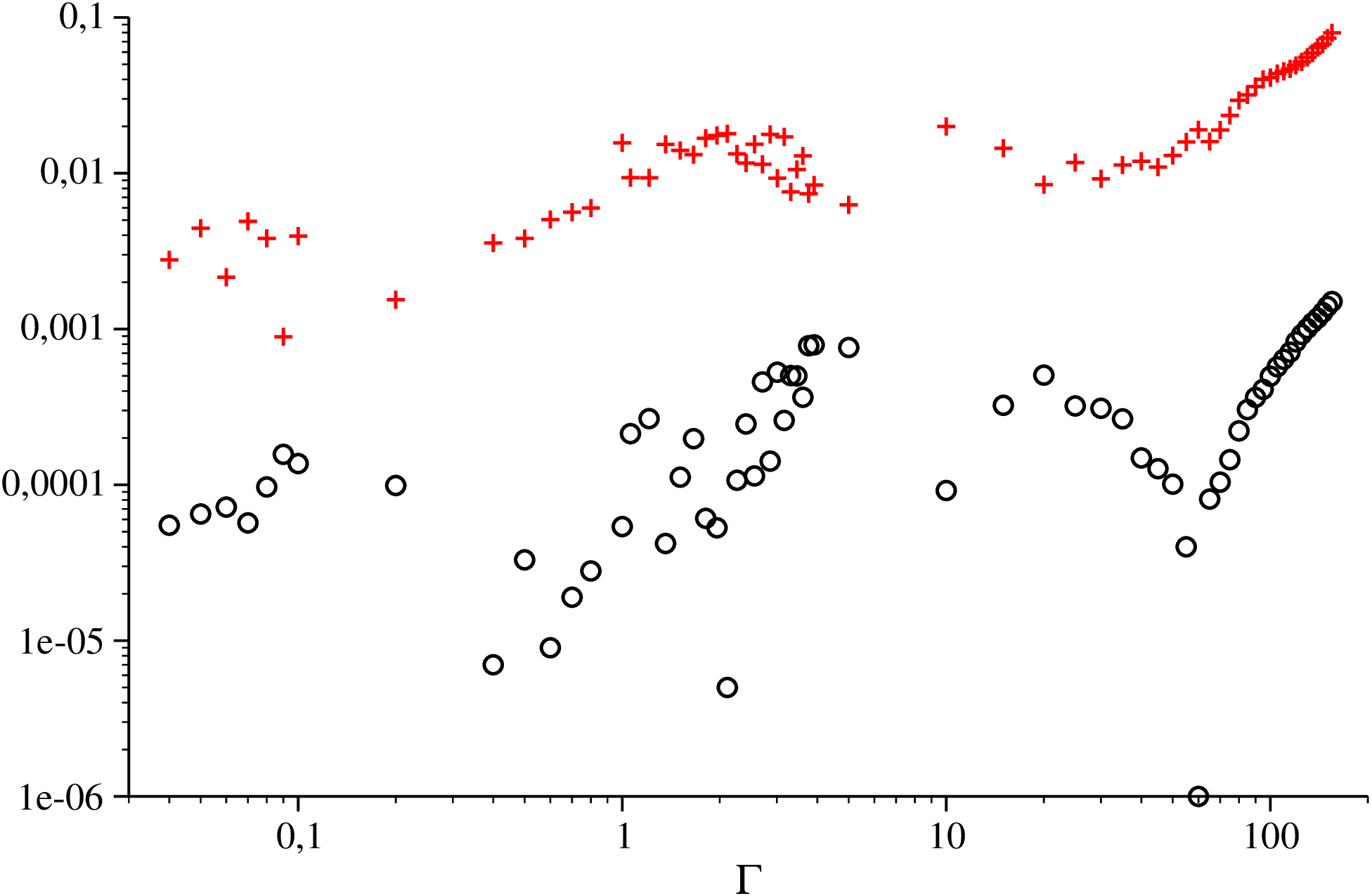}
\caption{Average (black circles) and maximum (red plus) absolute differences between fits of pair distribution function and simulation results for each studied value of the coupling parameter $\Gamma$.} 
\label{fig:ecarts_pdf}
\end{center}
\end{figure}

A comparison of the parametrized PDFs with the ones obtained in simulations for five values of $\Gamma$ spanning  weak to strong coupling is reported in Fig.\,\ref{fig:pdf}. It can be seen that the main features are nicely reproduced. The average and maximum discrepancies between the parametrization and the simulation  results as functions of $\Gamma$ are also reported in Fig. \ref{fig:ecarts_pdf}. The mean with respect to $\Gamma$ of the average and maximum discrepancies are $3.8\;10^{-4}$ and $2.\;10^{-2}$ respectively.

\begin{figure}[!t]
\begin{center}
\includegraphics[width=7.7cm]{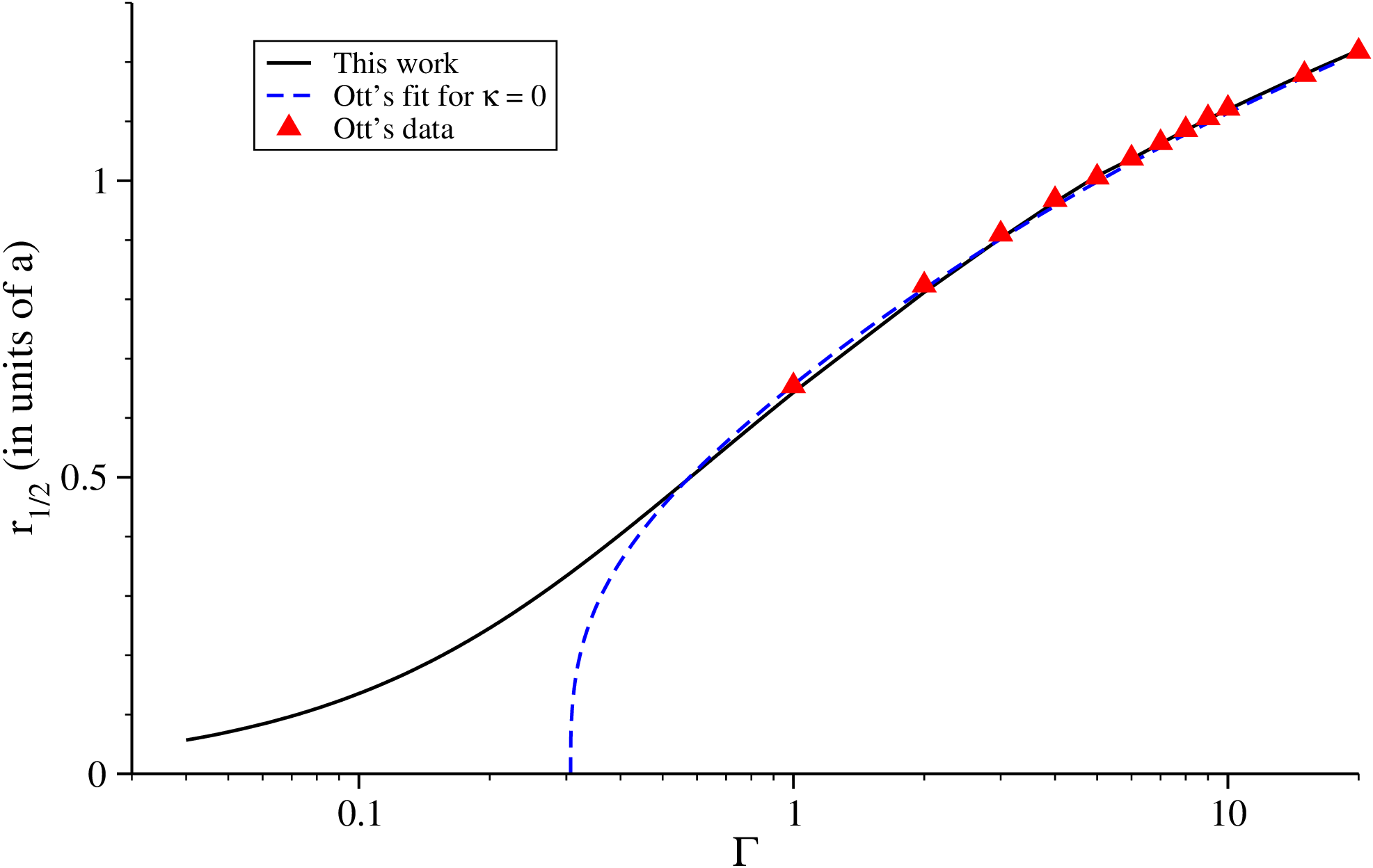}
\caption{$r_{1/2}$, the nearest distance to an ion where $g(r)=1/2$, as a function of $\Gamma$   in this work (black solid line) and compared to data (red triangles) and fit (blue dashed line) from Ott's work \cite{OttEtAl2014b,OttEtAl2015} (Ott's fit is extrapolated  for $\Gamma\le 1$).}
\label{fig:rhalf_lowG}
\end{center}
\end{figure}

We compared our parametrization with Ott \textit{et al.}'s results \cite{OttEtAl2014b,OttEtAl2015} for the height $g_\mathrm{max}$ of the first peak of the PDF and for the radius of the correlation hole $r_{1/2}$. The agreement is very good
 for $\Gamma\ge 1$ with an average discrepancy of 0.22\% for $r_{1/2}$ and 0.42\% for $g_\mathrm{max}$. 
However, Ott's fit cannot be extrapolated at lower values of $\Gamma$ in the case of the Coulomb system ($\kappa = 0$). Fig. \ref{fig:rhalf_lowG} shows how Ott's fit diverges from our results for $\Gamma \le 0.5$. The dependence of $r_{1/2}$ on $\Gamma$ and $\kappa$ deserves further investigation at low coupling.

\section{Static Structure Factor}
\label{SSF}

The SSF $S(k)$ is related to the Fourier transform of the correlation function, $h(r) = 1 - g(r)$, by the following relation:

\begin{equation}
S({\mathbf k}) = 1 + n\int d{\mathbf r}\, e^{-i{\mathbf k\cdot \mathbf r}} \left[ g(r) - 1 \right].
\end{equation}
In the reduced units introduced in Sec.\,\ref{MD}, an integration over angles leads to the quadrature to be used with our parametrization:

\begin{equation}
S_i(k) = 1 + 3\int_0^{\infty}  r^2 \left[ g(r) - 1 \right] \frac{\sin{(kr)}}{kr}\,dr.
\end{equation}
In the latter equation, and in the following, the wave number $k$ is in units of $a^{-1}$.

In order to prevent artificial oscillations occurring in the quadrature at low $k$ for high values of $\Gamma$ and to recover the low-$k$ behavior \cite{BausEtAl1980} ${k^2}/({k^2+3\Gamma})$, the SSF calculation  is slightly modified according to:

\begin{subequations}
\begin{equation}
\label{eq:ssf}
S_\mathrm{fit}(k) = 
\begin{cases}
S_{H}(k) & \text{if}~\Gamma \ge 60, \\
S_{L}(k) & \text{otherwise},
\end{cases}
\end{equation}
\begin{equation}
S_{H}(k) = 
\begin{cases}
S_i(k) & \text{if}~k \ge k_{=}, \\
S_\mathrm{high}(k) & \text{otherwise},
\end{cases}
\end{equation}
\begin{equation}
S_{L}(k) = S_i(k) \cdot W(k) + S_\mathrm{low}(k)\cdot (1-W(k)),
\end{equation}
\begin{equation}
S_\mathrm{high}(k) = \frac{S_m\left(\frac{k}{k_m}\right)^2}{1+\omega\cdot(k-k_m)^2},
\end{equation}
\begin{equation}
S_\mathrm{low}(k) = \frac{k^2}{k^2 + 3\Gamma},
\end{equation}
\begin{equation}
W(k) = \frac{1}{2}\left(1+\tanh{\left(\frac{k-k_l}{\delta k}\right)}\right),
\end{equation}
\end{subequations}
$k_{=}$ is defined as $k_{=} = \argmax_k{\left( S_\mathrm{high}(k) \ge S_i(k) \right)}$, the five remaining parameters are:
\begin{eqnarray}
S_m      & = & 1.211 + 1.079\:10^{-2}\;\Gamma, \nonumber \\
k_m      & = & 4.152 + 7.51\:10^{-4}\;\Gamma, \nonumber \\
\omega   & = & -0.978 + 4.84\:10^{-2}\;\Gamma, \\
k_l      & = & 1.25, \nonumber \\
\delta k & = & 0.25. \nonumber 
\end{eqnarray}

\begin{figure}[!t]
\begin{center}
\includegraphics[width=7.7cm]{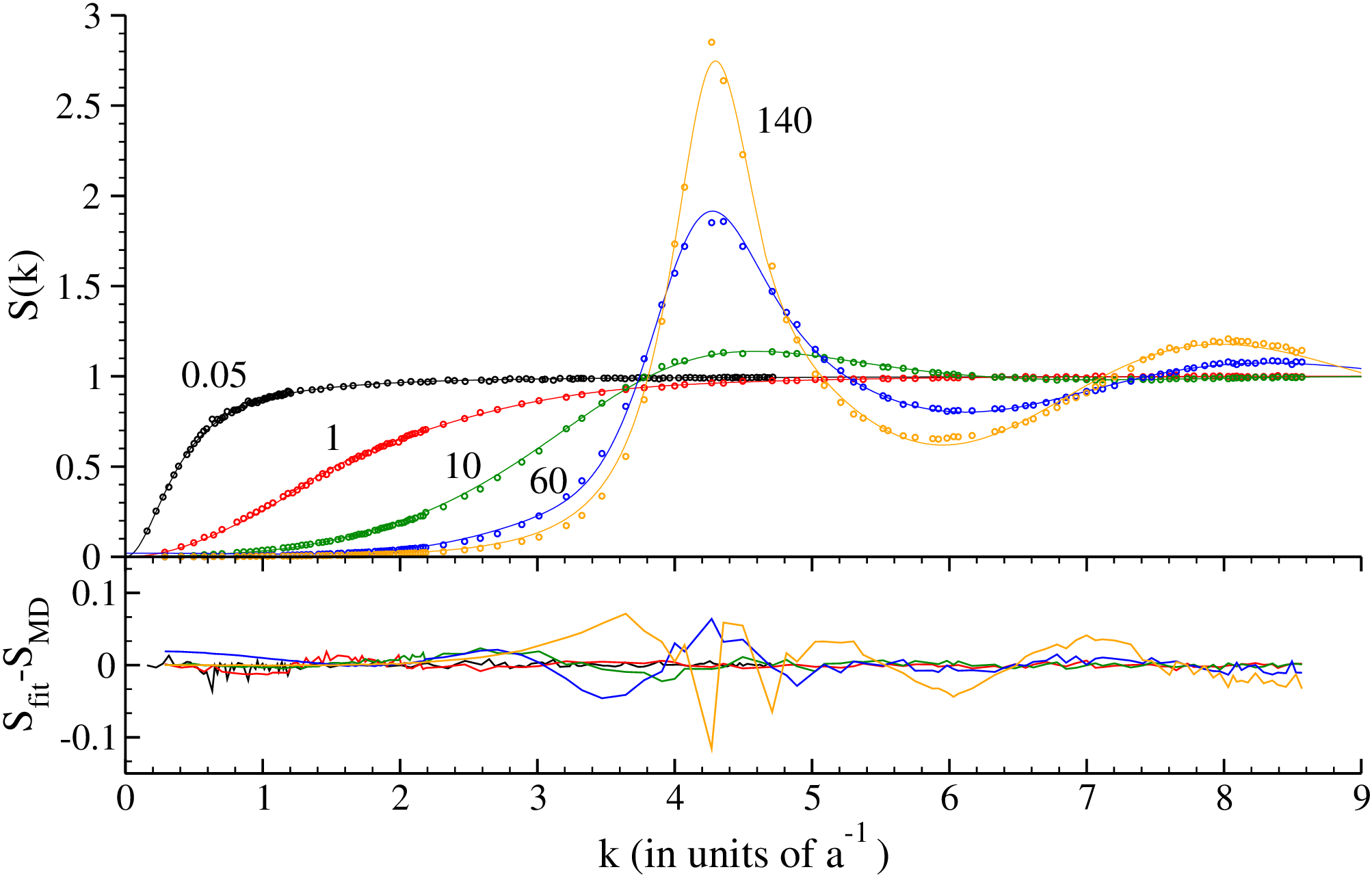}
\caption{Static structure factor for $\Gamma$ = 0.05 (black), 1 (red), 10 (green), 60 (blue) and 140 (orange). Top view: circles correspond to simulations while lines correspond to the parametrization. Bottom view: difference between fits ($S_\mathrm{fit}$) and simulations ($S_\mathrm{MD}$).} 
\label{fig:ssf}
\end{center}
\end{figure}

\begin{figure}[!t]
\begin{center}
\includegraphics[width=7.7cm]{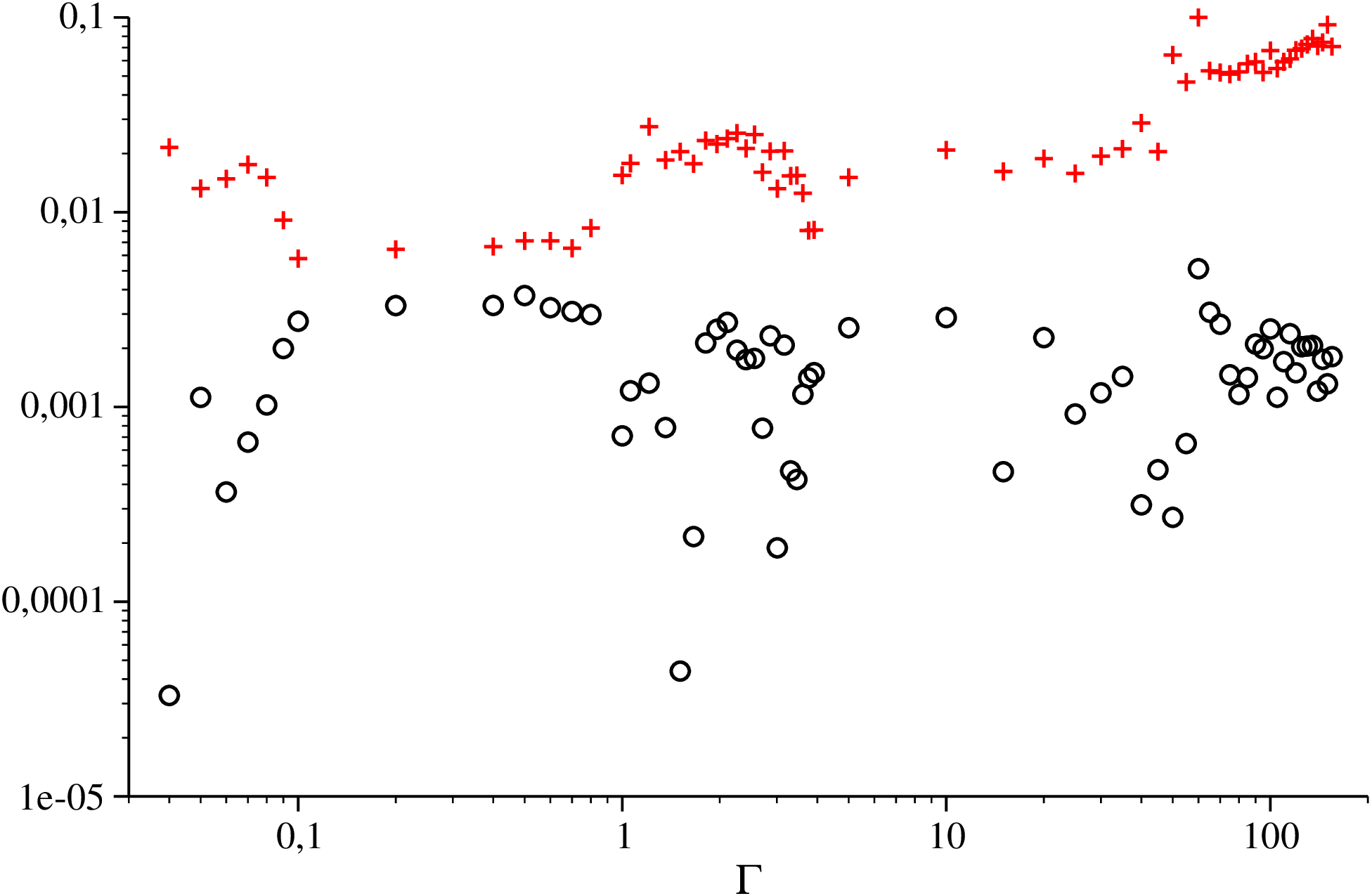}
\caption{Average (black circles) and maximum (red plus) absolute differences between fits of static structure factor and simulation results for each studied value of the coupling parameter $\Gamma$.} 
\label{fig:ecarts_ssf}
\end{center}
\end{figure}

The good agreement of the parametrized SSFs with the ones obtained in simulations is illustrated for five values of $\Gamma$ spanning weak to strong coupling in Fig. \ref{fig:ssf}. The average and maximum discrepancies between the parametrization and the simulation results as functions of $\Gamma$ are also reported in Fig. \ref{fig:ecarts_ssf}. The mean with respect to $\Gamma$ of the average and maximum discrepancies are $1.7\;10^{-3}$ and $3.3\;10^{-2}$.

The height $S_\mathrm{max}$ of the first peak of the SSF can be fitted as a function of $\Gamma$ by:

\begin{equation}
S_\mathrm{max} = 1 + 5.86\;10^{-3}\,\left[\ln{(\Gamma+1)}\right]^{3.56},
\end{equation}
with a maximum error of 0.039 and an average error of 1.5\;10$^{-4}$.

\section{Equation of state}
\label{EOS}

\begin{figure}[!t]
\begin{center}
\includegraphics[width=7.7cm]{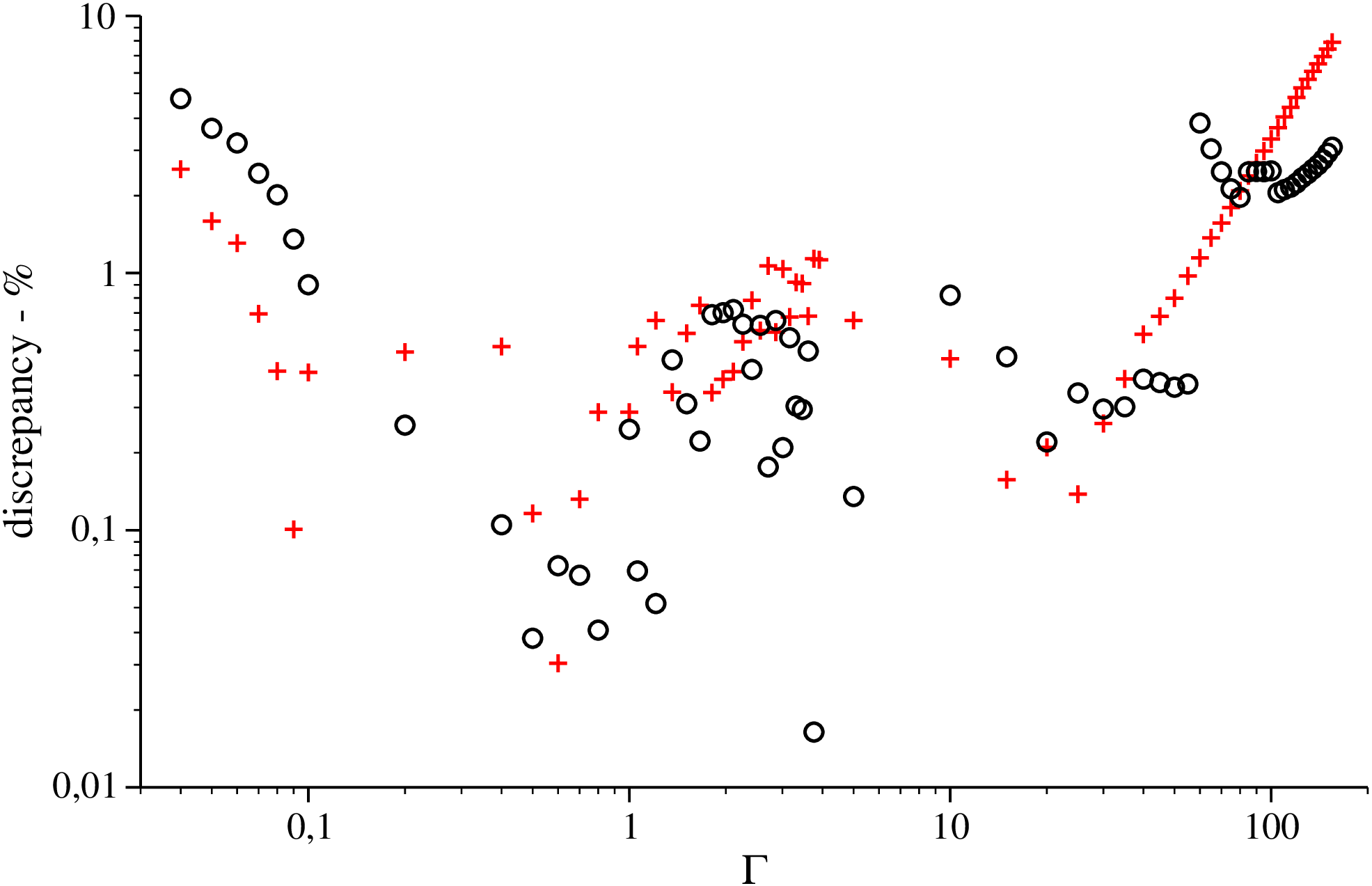}
\caption{Relative deviations in percent between reduced excess internal energy $U_{ex}$, per particle in unit of $k_B T\,\Gamma$, obtained in the simulations and computed using the parametrization of the pair distribution functions (red plus) or the static structure factors (black circles), for each studied value of the coupling parameter $\Gamma$.} 
\label{fig:pi_rules}
\end{center}
\end{figure}

We further assess the accuracy of the parametrization of the PDF and the SSF by computing the reduced excess internal energy $U_{ex}$ per particle in unit of $k_B T\,\Gamma$. It is related to the Coulomb interaction energy $U_C$, that is obtained directly from simulations, according to:
\begin{equation}
U_{ex} = -\frac{U_{C}}{N k_B T \, \Gamma}.
\end{equation}

This coulomb energy can also be computed from the PDF $g(r)$ or the SSF $S(k)$ with \cite{BausEtAl1980}
\begin{eqnarray}
\frac{U_{C}}{N} & = & \frac{n}{2} \int d{\mathbf r}\; U_C(r)\left[ g(r) - 1\right],\\
		& = & \frac{1}{2} \int \frac{d{\mathbf k}}{(2\pi)^3}\; U_C(k)\left[ S(k) - 1\right].
\end{eqnarray}
Using the reduced units introduced in Sec.\,\ref{MD}, these expressions simplify to:

\begin{equation}
\label{eq:uexcesg}
U_{ex} = -\frac{3}{2}\;\int_0^{\infty} r\left[ g(r) - 1\right] dr,
\end{equation}
for the PDF route and to:
\begin{equation}
\label{eq:uexcess}
U_{ex} = -\frac{1}{\pi}\;\int_0^{\infty} \left[ S(k) - 1\right] dk,
\end{equation}
for the SSF route. 
The Figure \ref{fig:pi_rules} shows the relative deviations between the excess energy $U_{ex}$ obtained from the simulation results and computed with the parametrization following the PDF-route (Eq. \ref{eq:uexcesg}) or the SSF-route (Eq. \ref{eq:uexcess}). In average, the relative accuracy is around 1-2\%. This gives a supplemental measurement of the quality of the fitted PDF and SSF proposed in this work.

\section{Yukawa Static Structure}
\label{YOCP}

The Yukawa static structure depends on two parameters: the Coulomb coupling parameter $\Gamma$ and the screening parameter $\kappa$. This renders any parametrization of the PDF and the SSF more complicated than in the case of the OCP static structure that only depends on $\Gamma$. A promising way to circumvent this difficulty was recently advanced by Ott \textit{et al.} \cite{OttEtAl2014b} with the definition of an effective coupling parameter   $\Gamma_\mathrm{eff}$  that establishes a correspondence between the PDFs of the OCP and Yukawa systems, at least at short-range. This effective coupling parameter $\Gamma_\mathrm{eff}$  was parametrized as a function of $\Gamma$ and $\kappa$ according to: \cite{OttEtAl2014b}
\begin{equation}
\label{gameff}
\Gamma_\mathrm{eff} = (1 - 0.309\, \kappa^2 + 0.0800\, \kappa^3)\, \Gamma,
\end{equation}
for $0 \leq \kappa \leq 2$ and $1 \leq  \Gamma_\mathrm{eff} \leq 150$.

\begin{figure}[!t]
\begin{center}
\includegraphics[width=7.7cm]{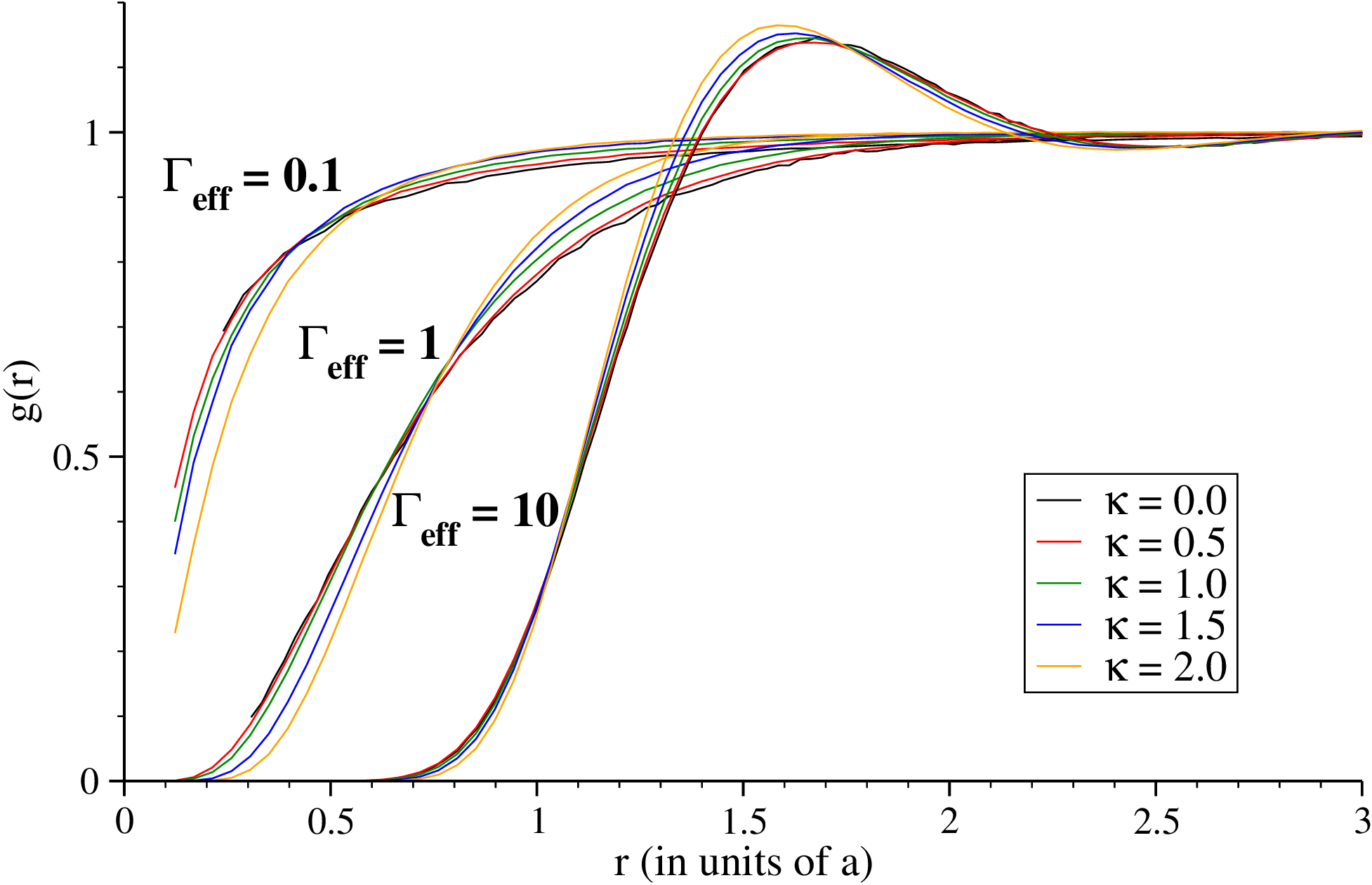}
\caption{Pair distribution functions of Coulomb ($\kappa$ = 0) and Yukawa ($\kappa$ = 0.5, 1.0, 1.5 and 2.0) systems corresponding to three values of the effective coupling parameter $\Gamma_\mathrm{eff}$ (0.1, 1 and 10), which is a function of $\Gamma$ and $\kappa$ given in Eq.\,\eqref{gameff}.} 
\label{fig:yukawapdf}
\end{center}
\end{figure}

The Figure \ref{fig:yukawapdf} shows that the PDFs of the Coulomb and Yukawa systems corresponding to the same value of the effective coupling parameter $\Gamma_\mathrm{eff}$ are indeed quite close to each others. However, there is still room for improvements especially at low coupling where the parametrization of $\Gamma_\mathrm{eff}$ is extrapolated. 

\begin{figure}[!t]
\begin{center}
\includegraphics[width=7.7cm]{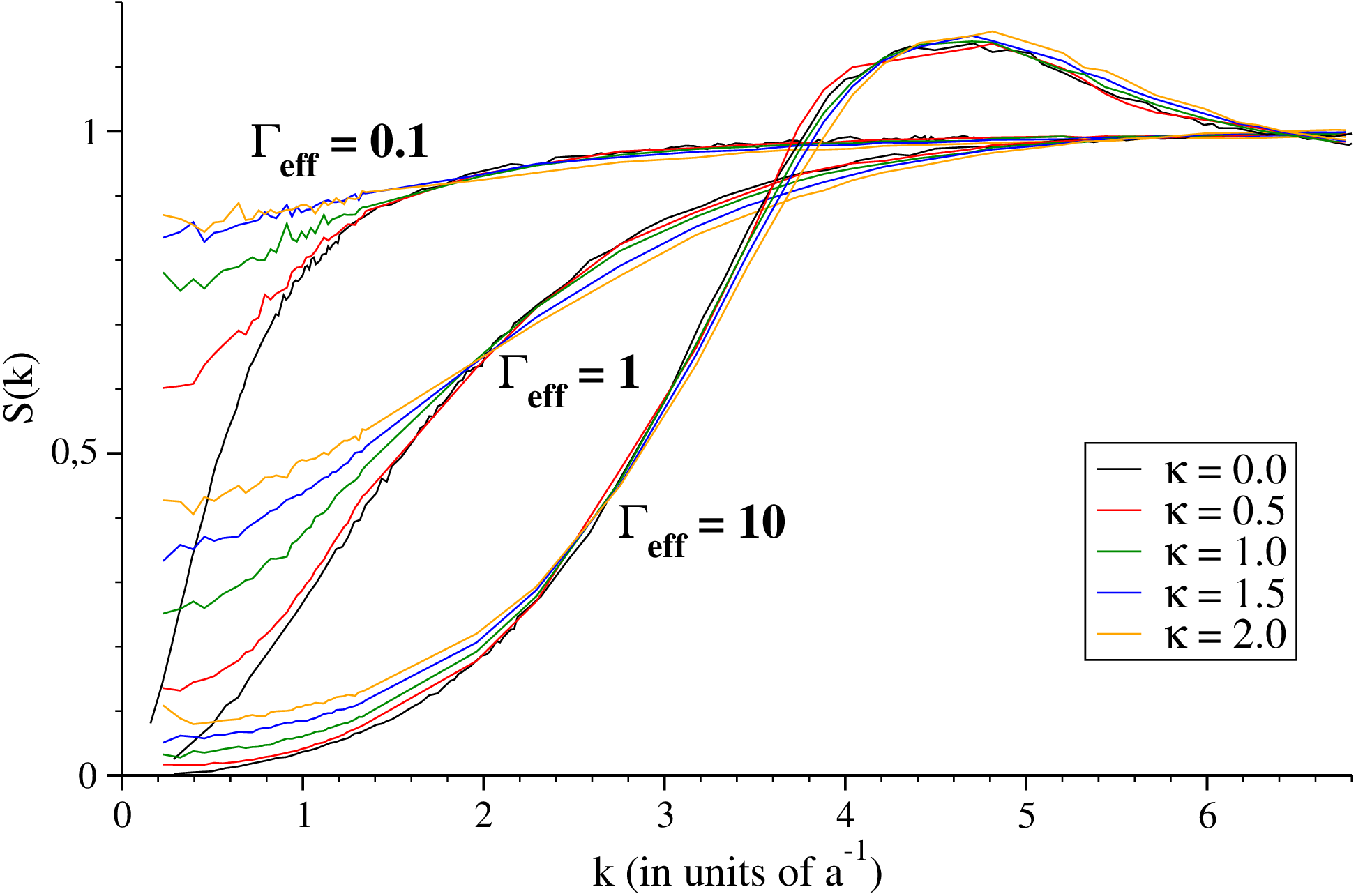}
\caption{Static structure factors of Coulomb ($\kappa$ = 0.) and Yukawa ($\kappa$ = 0.5, 1.0, 1.5 and 2.0) systems corresponding to three values of the effective coupling parameter $\Gamma_\mathrm{eff}$ (0.1, 1 and 10), which is a function of $\Gamma$ and $\kappa$ given in Eq.\,\eqref{gameff}.} 
\label{fig:yukawassf}
\end{center}
\end{figure}

Actually, this correspondence between the Coulomb and Yukawa systems is no longer possible for the long-range correlations, as revealed in the limit of vanishing wave number of the SSFs (see Figure \ref{fig:yukawassf}). This was expected since the value of the SSF in this limit is proportional to the compressibility of the system, \cite{HansenEtAl2006} which vanishes for the OCP but stays finite for the Yukawa systems. 

This fundamental difference between the OCP and Yukawa system renders impossible a perfect match between their static structure through the definition of an effective coupling parameter $\Gamma_\mathrm{eff}$. Nevertheless, it is still desirable to get the best approximate parametrization of $\Gamma_\mathrm{eff}$ as a function of $\Gamma$ and $\kappa$ from the viewpoint of a simple parametrization of the static structure of both systems. To this end, our choice to parametrize separately the short wave number range of the SSF should allow to pass from the OCP static structure to the Yukawa one.  Interestingly, the SSFs observed in quantum molecular dynamics simulations exhibit both the behavior of the Yukawa SSF at small wave number and the OCP behavior at higher wave number. \cite{ClerouinEtAl2016}

Because it is out of the scope of the present paper, we postpone the extension of our parametrization to the Yukawa system for a future study.

\section{x-ray diffraction interpretation}
\label{XRTS}

The role played by the OCP and Yukawa models as reference systems can be evidenced in the interpretation of experimental results. In this respect, the recent measurements of x-ray Thomson scattering \cite{GlenzerEtAl2009} gives unprecedented insights into the microscopic structure of WDM. 

Very recently, x-ray diffraction \cite{FletcherEtAl2015} was measured simultaneously with x-ray diffusion giving access to the static structure of aluminum ions together with the electronic plasmon spectrum. The frequency-resolved spectra give the density via the shift of the plasmon peak and the temperature via the ratio of intensities between the elastic scattering and the plasmon. The simultaneous measurement of the wave number-resolved spectrum allows for an independent check of the thermodynamic state producing the static structure factor.

The wave number-resolved spectrum is also known as the ion feature $W(k)$. It is related to the SSF $S(k)$ by: \cite{Chihara2000, GregoriEtAl2003,GlenzerEtAl2009}
\begin{equation}
W(k) = \left | f(k) + q(k) \right|^2\,S(k),
\end{equation}
where $f(k)$ and $q(k)$ are the form factors of bound and free electron densities at an ion. Both the form factors and the SSF, to be used in x-ray Thomson scattering interpretation,  are the subjects of active current research involving innovative developments in quantum molecular dynamics simulations. \cite{PlagemannEtAl2015,BaczewskiEtAl2016}

Here, we propose a simple strategy to tackle the problem of the interpretation of x-ray diffraction from WDM, prior to more involved treatments. The SSF is taken as the OCP one depending on density $\rho$ and temperature $T$ via the coupling parameter $\Gamma$ and the wave number $k$ in units of $a^{-1}$, i.e. a function of $\rho$. The form factor is parametrized using a simple exponential dependence: 
\begin{equation}
f(k) + q(k) = Z \exp (-\lambda k),
\end{equation} 
where $Z$ is the atomic number of the element.

\begin{figure}[!t]
\begin{center}
\includegraphics[width=8.5cm]{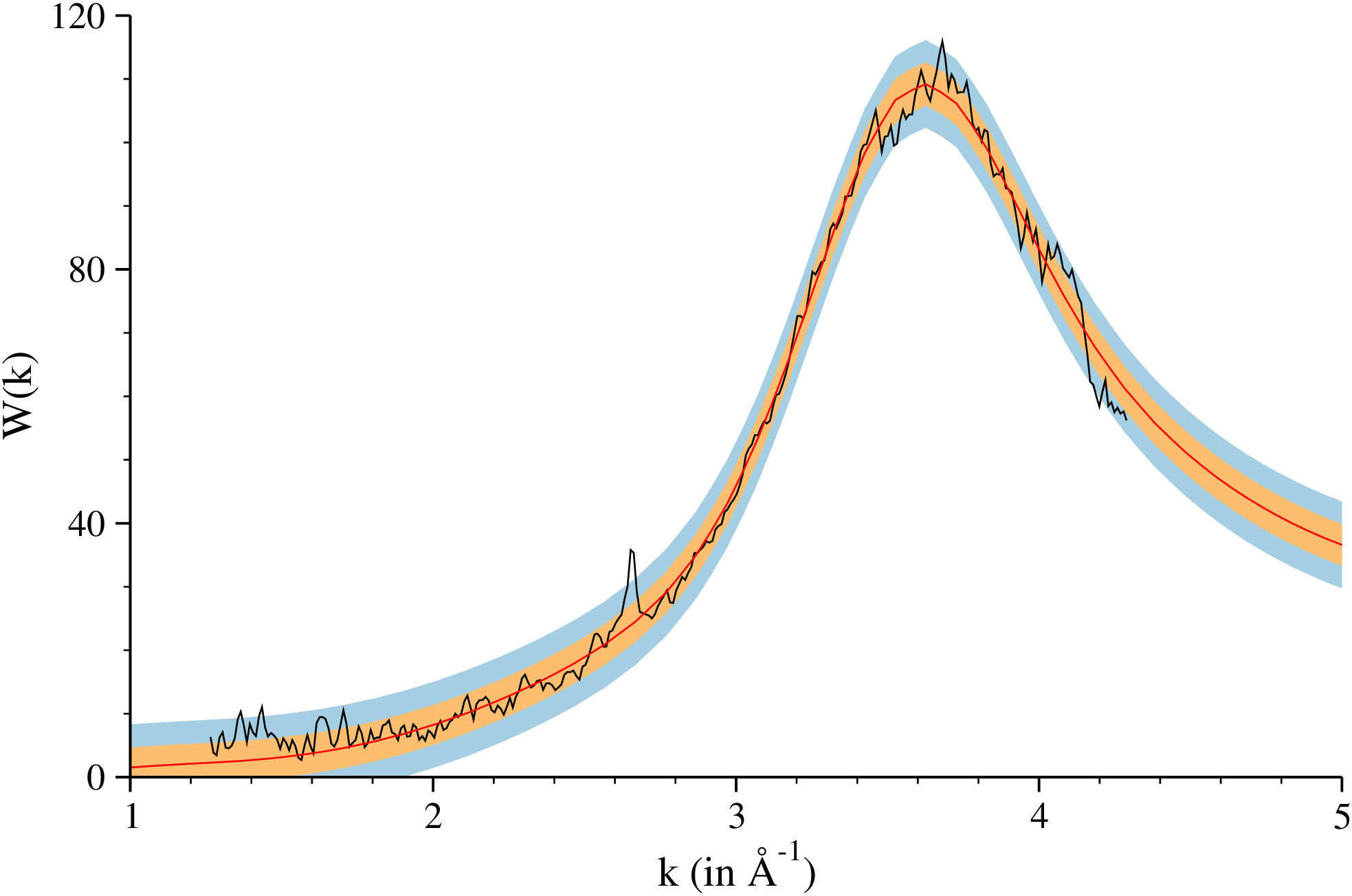}
\caption{Interpretation of the x-ray diffraction spectrum of shocked aluminum from Fletcher \textit{et al.} \cite{FletcherEtAl2015} (black line) with our OCP-based model and the best parameters $\hat{\beta} = \{\rho,T,\lambda\}$ given in Table\,\ref{tab:optim} (red line). Orange shaded area corresponds to the 95\% confidence interval coming from the model uncertainty ($\sigma_\mathrm{mod}$) and the parametric uncertainty on $\hat{\beta}$. Blue shaded area corresponds to the 95\% confidence interval coming from experimental uncertainty ($\sigma_\mathrm{exp}$), model uncertainty, and the parametric uncertainty.} 
\label{fig:ionfeature}
\end{center}
\end{figure}

\begin{table}[!b]
\centering
\begin{tabular}{cccc}
\hline 
$\beta$	&	$\hat\beta$	& $\sigma_{\hat\beta}$ 	& 	${\sigma_{\hat\beta}}/{\hat\beta}$ (\%)\\
\hline  
$\rho$ (g\,cm$^{-3}$)	& 6.813 	& 0.038   	&	0.6	\\
$T$ (eV)		& 2.25 	& 0.06 	&	2.7	\\
$\lambda$ (\AA)		& 0.138	& 0.001 &	0.7	\\
\hline 
$\sigma_\mathrm{mod}$		& 1.6	& 0.4	&	-	\\
\hline 
\end{tabular}
\caption{Best parameters $\hat{\beta}$ (mean and standard deviation) obtained by bayesian calibration and model uncertainty $\sigma_{mod}$.}
\label{tab:optim}
\end{table}

We illustrate the efficiency of this approach with the interpretation of the x-ray diffraction spectrum on aluminum measured by Fletcher \textit{et al.} \cite{FletcherEtAl2015} (see Fig.\,\ref{fig:ionfeature}).
We performed a Bayesian calibration \cite{FangEtAl2006,HastieEtAl2001,SantnerEtAl2003} of the three parameters $\beta=\{\rho,T,\lambda\}$, assuming the ionization $Q = 3$ close to the melting curve. Such a procedure requires many on-the-fly calculations of the SSF for different $\Gamma$ as presented above. 

This is not the scope of this paper to explain the Bayesian technique, so we only recall the few main points. We suppose that the $W(k)$ experimental data are subject to an uncertainty $\epsilon_\mathrm{exp}$ distributed according to a Gaussian distribution with a zero mean and a standard deviation $\sigma_\mathrm{exp}$ of 2.5 (this value is estimated from the noise affecting the measured spectrum in Fig.\,\ref{fig:ionfeature}). Our model is not perfect and hence reproduces the experimental data with an uncertainty $\epsilon_\mathrm{mod}$ distributed according to a Gaussian distribution with a zero mean and a standard deviation $\sigma_\mathrm{mod}$. The aim is to solve the Bayesian inference of the experimental dataset [$W(k)+\epsilon_\mathrm{exp}$] with our model [$W_\mathrm{mod}(k)+\epsilon_\mathrm{mod}$]. The Bayesian equation is sampled by use of a Markov Chain Monte Carlo (MCMC) procedure.\cite{FangEtAl2006,HastieEtAl2001,SantnerEtAl2003} 

This method provides three informations: 1) the best parameters $\hat{\beta}$ for our model, 2) the uncertainty on the $\hat{\beta}$ parameters and the correlations between them and 3) the model uncertainty $\sigma_\mathrm{mod}$, which is a measure of the quality of the model. These results are reported in Table \ref{tab:optim}. The uncertainty on the best parameters $\hat{\beta}$  are of the order of one percent. We checked that they are almost uncorrelated. The model uncertainty $\sigma_\mathrm{mod}$ is less than the experimental uncertainty $\sigma_\mathrm{exp}$, meaning that the model perfectly reproduces the experimental data. Indeed, the agreement between both experimental and theoretical ion features is excellent as appears in Fig.\,\ref{fig:ionfeature}. The simple exponential form factor with the best value of $\hat \lambda$ is consistent with different theoretical computations already published \cite{FletcherEtAl2015} (see also Ref.\,\citenum{ClerouinEtAl2016a}). 

The Bayesian method has allowed us to measure the accuracy of the OCP-based interpretation. However, a simple least-square fit method would provide roughly the same $\hat\beta$ parameters but without any information about the uncertainty on these $\hat\beta$ parameters nor any information about the global accuracy of the model.

Further confrontations with experimental results are still necessary to consolidate the status of the OCP model as a reference system for WDM. Some other examples of our OCP-based interpretation of x-ray Thomson scattering measurements can be found in Ref.\citenum{ClerouinEtAl2016a}.

\section{Conclusion}
\label{concl}

We have filled a gap in the corpus of rapidly available properties of the OCP model with the parametrization of its pair distribution function and its static structure factor. The whole fluid phase is covered from  weak to strong coupling. The accuracy of the fits was assessed by direct comparisons with molecular dynamics simulations and by calculations of the equation of state. 

Recent experiments on warm dense matter revealed the role of reference system the OCP model can play. This prompts us to provide a rapid evaluation of its static structure factor, since it is part of the ion feature in x-ray Thomson scattering measurements. As an illustration, we have used it to successfully interpret the recent x-ray diffraction experiment performed by Fletcher \textit{et al.} \cite{FletcherEtAl2015} to get the wave number-resolved spectrum of aluminum ion feature.

The extension of this work concerns the Yukawa model, more appropriate to dusty plasma. In recent experiments, \cite{BonitzEtAl2010} the pair distribution function is directly measured, and there is a pressing need to compare it with the Yukawa system. Ott \textit{et al.} \cite{OttEtAl2014b} suggested that an effective coupling parameter $\Gamma_\mathrm{eff}$ can be define as a function of the Coulomb coupling parameter $\Gamma$ and the screening parameter $\kappa$ to characterize the strength of the interaction in Yukawa system. It relies on the short-range structure of the plasma and establishes a correspondence between the OCP and the Yukawa pair distribution functions. We have shown that indeed using the fit of $\Gamma_\mathrm{eff}$ proposed by  Ott \textit{et al.} \cite{OttEtAl2014b} one can use our parametrization of the OCP pair distribution function for the Yukawa system. Nevertheless, we have also emphasized the limitations of this correspondence, especially in the weakly coupled regime. 

We plan to improve the fit of  $\Gamma_\mathrm{eff}$ at weak coupling and to investigate how the OCP and Yukawa static structure factors can be related. This latter issue highlights a fundamental difference between both models, since the static structure factor at vanishing wave number is proportional to the compressibility which is zero for the OCP and stays finite for the Yukawa system.

\section*{Acknowledgments}
We thank Vincent Dubois for providing tools to perform the Bayesian analysis and for many valuable advices. 


\begin{thebibliography}{51}
\expandafter\ifx\csname natexlab\endcsname\relax\def\natexlab#1{#1}\fi
\expandafter\ifx\csname bibnamefont\endcsname\relax
  \def\bibnamefont#1{#1}\fi
\expandafter\ifx\csname bibfnamefont\endcsname\relax
  \def\bibfnamefont#1{#1}\fi
\expandafter\ifx\csname citenamefont\endcsname\relax
  \def\citenamefont#1{#1}\fi
\providecommand{\bibinfo}[2]{#2}
\providecommand{\eprint}[2][]{\url{#2}}

\bibitem[{\citenamefont{Baraffe et~al.}(2010)\citenamefont{Baraffe, Chabrier,
  and Barman}}]{BaraffeEtAl2010}
\bibinfo{author}{\bibfnamefont{I.}~\bibnamefont{Baraffe}},
  \bibinfo{author}{\bibfnamefont{G.}~\bibnamefont{Chabrier}}, \bibnamefont{and}
  \bibinfo{author}{\bibfnamefont{T.}~\bibnamefont{Barman}},
  \bibinfo{journal}{Reports on Progress in Physics}
  \textbf{\bibinfo{volume}{73}}, \bibinfo{pages}{016901}
  (\bibinfo{year}{2010}).

\bibitem[{\citenamefont{Koester and Chanmugam}(1990)}]{KoesterEtAl1990}
\bibinfo{author}{\bibfnamefont{D.}~\bibnamefont{Koester}} \bibnamefont{and}
  \bibinfo{author}{\bibfnamefont{G.}~\bibnamefont{Chanmugam}},
  \bibinfo{journal}{Reports on Progress in Physics}
  \textbf{\bibinfo{volume}{53}}, \bibinfo{pages}{837} (\bibinfo{year}{1990}).

\bibitem[{\citenamefont{Daligault and Gupta}(2009)}]{DaligaultEtAl2009}
\bibinfo{author}{\bibfnamefont{J.}~\bibnamefont{Daligault}} \bibnamefont{and}
  \bibinfo{author}{\bibfnamefont{S.}~\bibnamefont{Gupta}},
  \bibinfo{journal}{The Astrophysical Journal} \textbf{\bibinfo{volume}{703}},
  \bibinfo{pages}{994} (\bibinfo{year}{2009}).

\bibitem[{\citenamefont{Bonitz et~al.}(2010)\citenamefont{Bonitz, Henning, and
  Block}}]{BonitzEtAl2010}
\bibinfo{author}{\bibfnamefont{M.}~\bibnamefont{Bonitz}},
  \bibinfo{author}{\bibfnamefont{C.}~\bibnamefont{Henning}}, \bibnamefont{and}
  \bibinfo{author}{\bibfnamefont{D.}~\bibnamefont{Block}},
  \bibinfo{journal}{Reports on Progress in Physics}
  \textbf{\bibinfo{volume}{73}}, \bibinfo{pages}{066501}
  (\bibinfo{year}{2010}).

\bibitem[{\citenamefont{Graziani et~al.}(2014)\citenamefont{Graziani,
  Desjarlais, Redmer, and Trickey}}]{GrazianiEtAl2014}
\bibinfo{editor}{\bibfnamefont{F.~R.} \bibnamefont{Graziani}},
  \bibinfo{editor}{\bibfnamefont{M.~P.} \bibnamefont{Desjarlais}},
  \bibinfo{editor}{\bibfnamefont{R.}~\bibnamefont{Redmer}}, \bibnamefont{and}
  \bibinfo{editor}{\bibfnamefont{S.~B.} \bibnamefont{Trickey}}, eds.,
  \emph{\bibinfo{title}{Frontiers and Challenges in Warm Dense Matter}},
  vol.~\bibinfo{volume}{96} of \emph{\bibinfo{series}{Lecture Notes in
  Computational Science and Engineering}} (\bibinfo{publisher}{Springer
  International Publishing Springer International Publishing},
  \bibinfo{address}{Switzerland}, \bibinfo{year}{2014}).

\bibitem[{\citenamefont{Lindl et~al.}(2004)\citenamefont{Lindl, Amendt, Berger,
  Glendinning, Glenzer, Haan, Kauffman, Landen, and Suter}}]{LindlEtAl2004}
\bibinfo{author}{\bibfnamefont{J.~D.} \bibnamefont{Lindl}},
  \bibinfo{author}{\bibfnamefont{P.}~\bibnamefont{Amendt}},
  \bibinfo{author}{\bibfnamefont{R.~L.} \bibnamefont{Berger}},
  \bibinfo{author}{\bibfnamefont{S.~G.} \bibnamefont{Glendinning}},
  \bibinfo{author}{\bibfnamefont{S.~H.} \bibnamefont{Glenzer}},
  \bibinfo{author}{\bibfnamefont{S.~W.} \bibnamefont{Haan}},
  \bibinfo{author}{\bibfnamefont{R.~L.} \bibnamefont{Kauffman}},
  \bibinfo{author}{\bibfnamefont{O.~L.} \bibnamefont{Landen}},
  \bibnamefont{and} \bibinfo{author}{\bibfnamefont{L.~J.} \bibnamefont{Suter}},
  \bibinfo{journal}{Physics of Plasmas} \textbf{\bibinfo{volume}{11}},
  \bibinfo{pages}{339} (\bibinfo{year}{2004}).

\bibitem[{\citenamefont{Shimoji}(1977)}]{Shimoji1977}
\bibinfo{author}{\bibfnamefont{M.}~\bibnamefont{Shimoji}},
  \emph{\bibinfo{title}{{L}iquid metals}} (\bibinfo{publisher}{Academic press,
  London}, \bibinfo{year}{1977}).

\bibitem[{\citenamefont{Hansen}(1973)}]{Hansen1973}
\bibinfo{author}{\bibfnamefont{J.~P.} \bibnamefont{Hansen}},
  \bibinfo{journal}{Phys. Rev. A} \textbf{\bibinfo{volume}{8}},
  \bibinfo{pages}{3096} (\bibinfo{year}{1973}).

\bibitem[{\citenamefont{Baus and Hansen}(1980)}]{BausEtAl1980}
\bibinfo{author}{\bibfnamefont{M.}~\bibnamefont{Baus}} \bibnamefont{and}
  \bibinfo{author}{\bibfnamefont{J.-P.} \bibnamefont{Hansen}},
  \bibinfo{journal}{Physics Reports} \textbf{\bibinfo{volume}{59}},
  \bibinfo{pages}{1} (\bibinfo{year}{1980}), ISSN \bibinfo{issn}{0370-1573}.

\bibitem[{\citenamefont{Hamaguchi et~al.}(1997)\citenamefont{Hamaguchi,
  Farouki, and Dubin}}]{HamaguchiEtAl1997}
\bibinfo{author}{\bibfnamefont{S.}~\bibnamefont{Hamaguchi}},
  \bibinfo{author}{\bibfnamefont{R.~T.} \bibnamefont{Farouki}},
  \bibnamefont{and} \bibinfo{author}{\bibfnamefont{D.~H.~E.}
  \bibnamefont{Dubin}}, \bibinfo{journal}{Phys. Rev. E}
  \textbf{\bibinfo{volume}{56}}, \bibinfo{pages}{4671} (\bibinfo{year}{1997}).

\bibitem[{\citenamefont{Caillol}(1999)}]{Caillol1999}
\bibinfo{author}{\bibfnamefont{J.~M.} \bibnamefont{Caillol}},
  \bibinfo{journal}{The Journal of Chemical Physics}
  \textbf{\bibinfo{volume}{111}}, \bibinfo{pages}{6538} (\bibinfo{year}{1999}).

\bibitem[{\citenamefont{Caillol and Gilles}(2010)}]{CaillolEtAl2010}
\bibinfo{author}{\bibfnamefont{J.-M.} \bibnamefont{Caillol}} \bibnamefont{and}
  \bibinfo{author}{\bibfnamefont{D.}~\bibnamefont{Gilles}},
  \bibinfo{journal}{Journal of Physics A: Mathematical and Theoretical}
  \textbf{\bibinfo{volume}{43}}, \bibinfo{pages}{105501}
  (\bibinfo{year}{2010}).

\bibitem[{\citenamefont{Bastea}(2005)}]{Bastea2005}
\bibinfo{author}{\bibfnamefont{S.}~\bibnamefont{Bastea}},
  \bibinfo{journal}{Phys.~Rev.~E} \textbf{\bibinfo{volume}{71}},
  \bibinfo{pages}{056405} (\bibinfo{year}{2005}).

\bibitem[{\citenamefont{Daligault et~al.}(2014)\citenamefont{Daligault,
  Rasmussen, and Baalrud}}]{DaligaultEtAl2014}
\bibinfo{author}{\bibfnamefont{J.}~\bibnamefont{Daligault}},
  \bibinfo{author}{\bibfnamefont{K.~O.} \bibnamefont{Rasmussen}},
  \bibnamefont{and} \bibinfo{author}{\bibfnamefont{S.~D.}
  \bibnamefont{Baalrud}}, \bibinfo{journal}{Phys. Rev. E}
  \textbf{\bibinfo{volume}{90}}, \bibinfo{pages}{033105}
  (\bibinfo{year}{2014}).

\bibitem[{\citenamefont{Daligault}(2006)}]{Daligault2006}
\bibinfo{author}{\bibfnamefont{J.}~\bibnamefont{Daligault}},
  \bibinfo{journal}{Phys. Rev. Lett.} \textbf{\bibinfo{volume}{96}},
  \bibinfo{pages}{065003} (\bibinfo{year}{2006}).

\bibitem[{\citenamefont{Daligault}(2009)}]{Daligault2009}
\bibinfo{author}{\bibfnamefont{J.}~\bibnamefont{Daligault}},
  \bibinfo{journal}{Phys. Rev. Lett.} \textbf{\bibinfo{volume}{103}},
  \bibinfo{pages}{029901} (\bibinfo{year}{2009}).

\bibitem[{\citenamefont{Cl\'erouin et~al.}(2013)\citenamefont{Cl\'erouin,
  Robert, Arnault, Kress, and Collins}}]{ClerouinEtAl2013}
\bibinfo{author}{\bibfnamefont{J.}~\bibnamefont{Cl\'erouin}},
  \bibinfo{author}{\bibfnamefont{G.}~\bibnamefont{Robert}},
  \bibinfo{author}{\bibfnamefont{P.}~\bibnamefont{Arnault}},
  \bibinfo{author}{\bibfnamefont{J.~D.} \bibnamefont{Kress}}, \bibnamefont{and}
  \bibinfo{author}{\bibfnamefont{L.~A.} \bibnamefont{Collins}},
  \bibinfo{journal}{Phys. Rev. E} \textbf{\bibinfo{volume}{87}},
  \bibinfo{pages}{061101} (\bibinfo{year}{2013}).

\bibitem[{\citenamefont{Arnault et~al.}(2013)\citenamefont{Arnault,
  {Cl{\'e}rouin}, Robert, Ticknor, Kress, and Collins}}]{ArnaultEtAl2013}
\bibinfo{author}{\bibfnamefont{P.}~\bibnamefont{Arnault}},
  \bibinfo{author}{\bibfnamefont{J.}~\bibnamefont{{Cl{\'e}rouin}}},
  \bibinfo{author}{\bibfnamefont{G.}~\bibnamefont{Robert}},
  \bibinfo{author}{\bibfnamefont{C.}~\bibnamefont{Ticknor}},
  \bibinfo{author}{\bibfnamefont{J.~D.} \bibnamefont{Kress}}, \bibnamefont{and}
  \bibinfo{author}{\bibfnamefont{L.~A.} \bibnamefont{Collins}},
  \bibinfo{journal}{Phys. Rev. E} \textbf{\bibinfo{volume}{88}},
  \bibinfo{pages}{063106} (\bibinfo{year}{2013}).

\bibitem[{\citenamefont{{Cl\'erouin}}(2015)}]{Clerouin2015}
\bibinfo{author}{\bibfnamefont{J.}~\bibnamefont{{Cl\'erouin}}},
  \bibinfo{journal}{Molecular Physics} \textbf{\bibinfo{volume}{113}},
  \bibinfo{pages}{2403} (\bibinfo{year}{2015}).

\bibitem[{\citenamefont{Cl\'erouin et~al.}(2015)\citenamefont{Cl\'erouin,
  Robert, Arnault, Ticknor, Kress, and Collins}}]{ClerouinEtAl2015a}
\bibinfo{author}{\bibfnamefont{J.}~\bibnamefont{Cl\'erouin}},
  \bibinfo{author}{\bibfnamefont{G.}~\bibnamefont{Robert}},
  \bibinfo{author}{\bibfnamefont{P.}~\bibnamefont{Arnault}},
  \bibinfo{author}{\bibfnamefont{C.}~\bibnamefont{Ticknor}},
  \bibinfo{author}{\bibfnamefont{J.~D.} \bibnamefont{Kress}}, \bibnamefont{and}
  \bibinfo{author}{\bibfnamefont{L.~A.} \bibnamefont{Collins}},
  \bibinfo{journal}{Phys. Rev. E.} \textbf{\bibinfo{volume}{91}},
  \bibinfo{pages}{011101(R)} (\bibinfo{year}{2015}).

\bibitem[{\citenamefont{Arnault}(2013)}]{Arnault2013}
\bibinfo{author}{\bibfnamefont{P.}~\bibnamefont{Arnault}},
  \bibinfo{journal}{High Energy Density Physics} \textbf{\bibinfo{volume}{9}},
  \bibinfo{pages}{711} (\bibinfo{year}{2013}).

\bibitem[{\citenamefont{Whitley et~al.}(2015)\citenamefont{Whitley, Alley,
  Cabot, Castor, Nilsen, and DeWitt}}]{WhitleyEtAl2015}
\bibinfo{author}{\bibfnamefont{H.~D.} \bibnamefont{Whitley}},
  \bibinfo{author}{\bibfnamefont{W.~E.} \bibnamefont{Alley}},
  \bibinfo{author}{\bibfnamefont{W.~H.} \bibnamefont{Cabot}},
  \bibinfo{author}{\bibfnamefont{J.~I.} \bibnamefont{Castor}},
  \bibinfo{author}{\bibfnamefont{J.}~\bibnamefont{Nilsen}}, \bibnamefont{and}
  \bibinfo{author}{\bibfnamefont{H.~E.} \bibnamefont{DeWitt}},
  \bibinfo{journal}{Contributions to Plasma Physics}
  \textbf{\bibinfo{volume}{55}}, \bibinfo{pages}{413} (\bibinfo{year}{2015}),
  ISSN \bibinfo{issn}{1521-3986}.

\bibitem[{\citenamefont{Ticknor et~al.}(2016)\citenamefont{Ticknor, Kress,
  Collins, Cl{\'e}rouin, Arnault, and Decoster}}]{TicknorEtAl2016}
\bibinfo{author}{\bibfnamefont{C.}~\bibnamefont{Ticknor}},
  \bibinfo{author}{\bibfnamefont{J.~D.} \bibnamefont{Kress}},
  \bibinfo{author}{\bibfnamefont{L.~A.} \bibnamefont{Collins}},
  \bibinfo{author}{\bibfnamefont{J.}~\bibnamefont{Cl{\'e}rouin}},
  \bibinfo{author}{\bibfnamefont{P.}~\bibnamefont{Arnault}}, \bibnamefont{and}
  \bibinfo{author}{\bibfnamefont{A.}~\bibnamefont{Decoster}},
  \bibinfo{journal}{Phys. Rev. E} \textbf{\bibinfo{volume}{Accepted}}
  (\bibinfo{year}{2016}).

\bibitem[{\citenamefont{Stanton and Murillo}(2015)}]{StantonEtAl2015}
\bibinfo{author}{\bibfnamefont{L.~G.} \bibnamefont{Stanton}} \bibnamefont{and}
  \bibinfo{author}{\bibfnamefont{M.~S.} \bibnamefont{Murillo}},
  \bibinfo{journal}{Phys. Rev. E} \textbf{\bibinfo{volume}{91}},
  \bibinfo{pages}{033104} (\bibinfo{year}{2015}).

\bibitem[{\citenamefont{Hansen and McDonald}(2006)}]{HansenEtAl2006}
\bibinfo{author}{\bibfnamefont{J.-P.} \bibnamefont{Hansen}} \bibnamefont{and}
  \bibinfo{author}{\bibfnamefont{I.~R.} \bibnamefont{McDonald}},
  \emph{\bibinfo{title}{Theory of simple liquids}}
  (\bibinfo{publisher}{Academic Press Cambridge}, \bibinfo{year}{2006}),
  \bibinfo{edition}{3rd} ed.

\bibitem[{\citenamefont{Barker and Henderson}(1967)}]{BarkerEtAl1967}
\bibinfo{author}{\bibfnamefont{J.~A.} \bibnamefont{Barker}} \bibnamefont{and}
  \bibinfo{author}{\bibfnamefont{D.}~\bibnamefont{Henderson}},
  \bibinfo{journal}{The Journal of Chemical Physics}
  \textbf{\bibinfo{volume}{47}} (\bibinfo{year}{1967}).

\bibitem[{\citenamefont{Weeks et~al.}(1971)\citenamefont{Weeks, Chandler, and
  Andersen}}]{WeeksEtAl1971}
\bibinfo{author}{\bibfnamefont{J.~D.} \bibnamefont{Weeks}},
  \bibinfo{author}{\bibfnamefont{D.}~\bibnamefont{Chandler}}, \bibnamefont{and}
  \bibinfo{author}{\bibfnamefont{H.~C.} \bibnamefont{Andersen}},
  \bibinfo{journal}{The Journal of Chemical Physics}
  \textbf{\bibinfo{volume}{54}} (\bibinfo{year}{1971}).

\bibitem[{\citenamefont{Golden and Kalman}(2000)}]{GoldenEtAl2000}
\bibinfo{author}{\bibfnamefont{K.~I.} \bibnamefont{Golden}} \bibnamefont{and}
  \bibinfo{author}{\bibfnamefont{G.~J.} \bibnamefont{Kalman}},
  \bibinfo{journal}{Physics of Plasmas} \textbf{\bibinfo{volume}{7}},
  \bibinfo{pages}{14} (\bibinfo{year}{2000}).

\bibitem[{\citenamefont{Golden and Kalman}(2001)}]{GoldenEtAl2001}
\bibinfo{author}{\bibfnamefont{K.~I.} \bibnamefont{Golden}} \bibnamefont{and}
  \bibinfo{author}{\bibfnamefont{G.~J.} \bibnamefont{Kalman}},
  \bibinfo{journal}{Physics of Plasmas} \textbf{\bibinfo{volume}{8}},
  \bibinfo{pages}{5064} (\bibinfo{year}{2001}).

\bibitem[{\citenamefont{Rosenberg and Kalman}(1997)}]{RosenbergEtAl1997}
\bibinfo{author}{\bibfnamefont{M.}~\bibnamefont{Rosenberg}} \bibnamefont{and}
  \bibinfo{author}{\bibfnamefont{G.}~\bibnamefont{Kalman}},
  \bibinfo{journal}{Phys. Rev. E} \textbf{\bibinfo{volume}{56}},
  \bibinfo{pages}{7166} (\bibinfo{year}{1997}).

\bibitem[{\citenamefont{Chaturvedi et~al.}(1981)\citenamefont{Chaturvedi,
  Senatore, and Tosi}}]{ChaturvediEtAl1981}
\bibinfo{author}{\bibfnamefont{D.~K.} \bibnamefont{Chaturvedi}},
  \bibinfo{author}{\bibfnamefont{G.}~\bibnamefont{Senatore}}, \bibnamefont{and}
  \bibinfo{author}{\bibfnamefont{M.~P.} \bibnamefont{Tosi}},
  \bibinfo{journal}{Il Nuovo Cimento B (1971-1996)}
  \textbf{\bibinfo{volume}{62}}, \bibinfo{pages}{375} (\bibinfo{year}{1981}),
  ISSN \bibinfo{issn}{1826-9877}.

\bibitem[{\citenamefont{Rogers et~al.}(1983)\citenamefont{Rogers, Young,
  DeWitt, and Ross}}]{RogersEtAl1983}
\bibinfo{author}{\bibfnamefont{F.~J.} \bibnamefont{Rogers}},
  \bibinfo{author}{\bibfnamefont{D.~A.} \bibnamefont{Young}},
  \bibinfo{author}{\bibfnamefont{H.~E.} \bibnamefont{DeWitt}},
  \bibnamefont{and} \bibinfo{author}{\bibfnamefont{M.}~\bibnamefont{Ross}},
  \bibinfo{journal}{Phys. Rev. A} \textbf{\bibinfo{volume}{28}},
  \bibinfo{pages}{2990} (\bibinfo{year}{1983}).

\bibitem[{\citenamefont{Ott et~al.}(2014)\citenamefont{Ott, Bonitz, Stanton,
  and Murillo}}]{OttEtAl2014b}
\bibinfo{author}{\bibfnamefont{T.}~\bibnamefont{Ott}},
  \bibinfo{author}{\bibfnamefont{M.}~\bibnamefont{Bonitz}},
  \bibinfo{author}{\bibfnamefont{L.~G.} \bibnamefont{Stanton}},
  \bibnamefont{and} \bibinfo{author}{\bibfnamefont{M.~S.}
  \bibnamefont{Murillo}}, \bibinfo{journal}{Physics of Plasmas}
  \textbf{\bibinfo{volume}{21}}, \bibinfo{pages}{113704}
  (\bibinfo{year}{2014}).

\bibitem[{\citenamefont{Ott and Bonitz}(2015)}]{OttEtAl2015}
\bibinfo{author}{\bibfnamefont{T.}~\bibnamefont{Ott}} \bibnamefont{and}
  \bibinfo{author}{\bibfnamefont{M.}~\bibnamefont{Bonitz}},
  \bibinfo{journal}{Contributions to Plasma Physics}
  \textbf{\bibinfo{volume}{55}}, \bibinfo{pages}{243} (\bibinfo{year}{2015}),
  ISSN \bibinfo{issn}{1521-3986}.

\bibitem[{\citenamefont{Limbach et~al.}(2006)\citenamefont{Limbach, Arnold,
  Mann, and Holm}}]{Limbach2006704}
\bibinfo{author}{\bibfnamefont{H.}~\bibnamefont{Limbach}},
  \bibinfo{author}{\bibfnamefont{A.}~\bibnamefont{Arnold}},
  \bibinfo{author}{\bibfnamefont{B.}~\bibnamefont{Mann}}, \bibnamefont{and}
  \bibinfo{author}{\bibfnamefont{C.}~\bibnamefont{Holm}},
  \bibinfo{journal}{Computer Physics Communications}
  \textbf{\bibinfo{volume}{174}}, \bibinfo{pages}{704 } (\bibinfo{year}{2006}),
  ISSN \bibinfo{issn}{0010-4655}.

\bibitem[{\citenamefont{Frenkel and Smit}(2002)}]{FrenkelEtAl2002}
\bibinfo{author}{\bibfnamefont{D.}~\bibnamefont{Frenkel}} \bibnamefont{and}
  \bibinfo{author}{\bibfnamefont{B.}~\bibnamefont{Smit}},
  \emph{\bibinfo{title}{Understanding Molecular Simulation}}
  (\bibinfo{publisher}{Academic Press}, \bibinfo{address}{San Diego},
  \bibinfo{year}{2002}), \bibinfo{edition}{2nd} ed.

\bibitem[{\citenamefont{Berendsen et~al.}(1984)\citenamefont{Berendsen, Postma,
  van Gunsteren, DiNola, and Haak}}]{BerendsenEtAl1984}
\bibinfo{author}{\bibfnamefont{H.~J.~C.} \bibnamefont{Berendsen}},
  \bibinfo{author}{\bibfnamefont{J.~P.~M.} \bibnamefont{Postma}},
  \bibinfo{author}{\bibfnamefont{W.~F.} \bibnamefont{van Gunsteren}},
  \bibinfo{author}{\bibfnamefont{A.}~\bibnamefont{DiNola}}, \bibnamefont{and}
  \bibinfo{author}{\bibfnamefont{J.~R.} \bibnamefont{Haak}},
  \bibinfo{journal}{J. Chem. Phys.} \textbf{\bibinfo{volume}{81}},
  \bibinfo{pages}{3684} (\bibinfo{year}{1984}).

\bibitem[{\citenamefont{Hockney and Eastwood}(1988)}]{HockneyEtAl1988}
\bibinfo{author}{\bibfnamefont{R.~W.} \bibnamefont{Hockney}} \bibnamefont{and}
  \bibinfo{author}{\bibfnamefont{J.~W.} \bibnamefont{Eastwood}},
  \emph{\bibinfo{title}{Computer Simulation Using Particles}}
  (\bibinfo{publisher}{IOP}, \bibinfo{year}{1988}).

\bibitem[{\citenamefont{Matteoli and Mansoori}(1995)}]{MatteoliEtAl1995}
\bibinfo{author}{\bibfnamefont{E.}~\bibnamefont{Matteoli}} \bibnamefont{and}
  \bibinfo{author}{\bibfnamefont{G.~A.} \bibnamefont{Mansoori}},
  \bibinfo{journal}{The Journal of Chemical Physics}
  \textbf{\bibinfo{volume}{103}}, \bibinfo{pages}{4672} (\bibinfo{year}{1995}).

\bibitem[{\citenamefont{Lai et~al.}(2012)\citenamefont{Lai, Hsieh, and
  Lin}}]{LaiEtAl2012}
\bibinfo{author}{\bibfnamefont{P.-K.} \bibnamefont{Lai}},
  \bibinfo{author}{\bibfnamefont{C.-H.} \bibnamefont{Hsieh}}, \bibnamefont{and}
  \bibinfo{author}{\bibfnamefont{S.-T.} \bibnamefont{Lin}},
  \bibinfo{journal}{Phys. Chem. Chem. Phys.} \textbf{\bibinfo{volume}{14}},
  \bibinfo{pages}{15206} (\bibinfo{year}{2012}).

\bibitem[{\citenamefont{Cl\'erouin et~al.}(2016)\citenamefont{Cl\'erouin,
  Arnault, Ticknor, Kress, and Collins}}]{ClerouinEtAl2016}
\bibinfo{author}{\bibfnamefont{J.}~\bibnamefont{Cl\'erouin}},
  \bibinfo{author}{\bibfnamefont{P.}~\bibnamefont{Arnault}},
  \bibinfo{author}{\bibfnamefont{C.}~\bibnamefont{Ticknor}},
  \bibinfo{author}{\bibfnamefont{J.~D.} \bibnamefont{Kress}}, \bibnamefont{and}
  \bibinfo{author}{\bibfnamefont{L.~A.} \bibnamefont{Collins}},
  \bibinfo{journal}{Phys. Rev. Lett.} \textbf{\bibinfo{volume}{116}},
  \bibinfo{pages}{115003} (\bibinfo{year}{2016}).

\bibitem[{\citenamefont{Glenzer and Redmer}(2009)}]{GlenzerEtAl2009}
\bibinfo{author}{\bibfnamefont{S.~H.} \bibnamefont{Glenzer}} \bibnamefont{and}
  \bibinfo{author}{\bibfnamefont{R.}~\bibnamefont{Redmer}},
  \bibinfo{journal}{Rev. Mod. Phys.} \textbf{\bibinfo{volume}{81}},
  \bibinfo{pages}{1625} (\bibinfo{year}{2009}).

\bibitem[{\citenamefont{Fletcher et~al.}(2015)\citenamefont{Fletcher, Lee,
  D{\"o}ppner, Galtier, Nagler, Heimann, Fortmann, LePape, Ma, Millot
  et~al.}}]{FletcherEtAl2015}
\bibinfo{author}{\bibfnamefont{L.~B.} \bibnamefont{Fletcher}},
  \bibinfo{author}{\bibfnamefont{H.~J.} \bibnamefont{Lee}},
  \bibinfo{author}{\bibfnamefont{T.}~\bibnamefont{D{\"o}ppner}},
  \bibinfo{author}{\bibfnamefont{E.}~\bibnamefont{Galtier}},
  \bibinfo{author}{\bibfnamefont{B.}~\bibnamefont{Nagler}},
  \bibinfo{author}{\bibfnamefont{P.}~\bibnamefont{Heimann}},
  \bibinfo{author}{\bibfnamefont{C.}~\bibnamefont{Fortmann}},
  \bibinfo{author}{\bibfnamefont{S.}~\bibnamefont{LePape}},
  \bibinfo{author}{\bibfnamefont{T.}~\bibnamefont{Ma}},
  \bibinfo{author}{\bibfnamefont{M.}~\bibnamefont{Millot}},
  \bibnamefont{et~al.}, \bibinfo{journal}{Nat Photon}
  \textbf{\bibinfo{volume}{9}}, \bibinfo{pages}{274} (\bibinfo{year}{2015}).

\bibitem[{\citenamefont{Chihara}(2000)}]{Chihara2000}
\bibinfo{author}{\bibfnamefont{J.}~\bibnamefont{Chihara}},
  \bibinfo{journal}{Journal of Physics: Condensed Matter}
  \textbf{\bibinfo{volume}{12}}, \bibinfo{pages}{231} (\bibinfo{year}{2000}).

\bibitem[{\citenamefont{Gregori et~al.}(2003)\citenamefont{Gregori, Glenzer,
  Rozmus, Lee, and Landen}}]{GregoriEtAl2003}
\bibinfo{author}{\bibfnamefont{G.}~\bibnamefont{Gregori}},
  \bibinfo{author}{\bibfnamefont{S.~H.} \bibnamefont{Glenzer}},
  \bibinfo{author}{\bibfnamefont{W.}~\bibnamefont{Rozmus}},
  \bibinfo{author}{\bibfnamefont{R.~W.} \bibnamefont{Lee}}, \bibnamefont{and}
  \bibinfo{author}{\bibfnamefont{O.~L.} \bibnamefont{Landen}},
  \bibinfo{journal}{Phys. Rev. E} \textbf{\bibinfo{volume}{67}},
  \bibinfo{pages}{026412} (\bibinfo{year}{2003}).

\bibitem[{\citenamefont{Plagemann et~al.}(2015)\citenamefont{Plagemann,
  R\"uter, Bornath, Shihab, Desjarlais, Fortmann, Glenzer, and
  Redmer}}]{PlagemannEtAl2015}
\bibinfo{author}{\bibfnamefont{K.-U.} \bibnamefont{Plagemann}},
  \bibinfo{author}{\bibfnamefont{H.~R.} \bibnamefont{R\"uter}},
  \bibinfo{author}{\bibfnamefont{T.}~\bibnamefont{Bornath}},
  \bibinfo{author}{\bibfnamefont{M.}~\bibnamefont{Shihab}},
  \bibinfo{author}{\bibfnamefont{M.~P.} \bibnamefont{Desjarlais}},
  \bibinfo{author}{\bibfnamefont{C.}~\bibnamefont{Fortmann}},
  \bibinfo{author}{\bibfnamefont{S.~H.} \bibnamefont{Glenzer}},
  \bibnamefont{and} \bibinfo{author}{\bibfnamefont{R.}~\bibnamefont{Redmer}},
  \bibinfo{journal}{Phys. Rev. E} \textbf{\bibinfo{volume}{92}},
  \bibinfo{pages}{013103} (\bibinfo{year}{2015}).

\bibitem[{\citenamefont{Baczewski et~al.}(2016)\citenamefont{Baczewski,
  Shulenburger, Desjarlais, Hansen, and Magyar}}]{BaczewskiEtAl2016}
\bibinfo{author}{\bibfnamefont{A.~D.} \bibnamefont{Baczewski}},
  \bibinfo{author}{\bibfnamefont{L.}~\bibnamefont{Shulenburger}},
  \bibinfo{author}{\bibfnamefont{M.~P.} \bibnamefont{Desjarlais}},
  \bibinfo{author}{\bibfnamefont{S.~B.} \bibnamefont{Hansen}},
  \bibnamefont{and} \bibinfo{author}{\bibfnamefont{R.~J.}
  \bibnamefont{Magyar}}, \bibinfo{journal}{Phys. Rev. Lett.}
  \textbf{\bibinfo{volume}{116}}, \bibinfo{pages}{115004}
  (\bibinfo{year}{2016}).

\bibitem[{\citenamefont{Fang et~al.}(2006)\citenamefont{Fang, Li, and
  Sudjianto}}]{FangEtAl2006}
\bibinfo{author}{\bibfnamefont{K.-T.} \bibnamefont{Fang}},
  \bibinfo{author}{\bibfnamefont{R.}~\bibnamefont{Li}}, \bibnamefont{and}
  \bibinfo{author}{\bibfnamefont{A.}~\bibnamefont{Sudjianto}},
  \emph{\bibinfo{title}{{D}esign and modeling for computer experiments}}
  (\bibinfo{publisher}{Chapman and Hall, Boca Raton}, \bibinfo{year}{2006}).

\bibitem[{\citenamefont{Hastie et~al.}(2001)\citenamefont{Hastie, Tibshirani,
  and Friedman}}]{HastieEtAl2001}
\bibinfo{author}{\bibfnamefont{T.}~\bibnamefont{Hastie}},
  \bibinfo{author}{\bibfnamefont{R.}~\bibnamefont{Tibshirani}},
  \bibnamefont{and} \bibinfo{author}{\bibfnamefont{J.}~\bibnamefont{Friedman}},
  \emph{\bibinfo{title}{{T}he elements of statistical learning: data mining,
  inference, and prediction}} (\bibinfo{publisher}{Springer, New York},
  \bibinfo{year}{2001}).

\bibitem[{\citenamefont{Santner et~al.}(2003)\citenamefont{Santner, Williams,
  and Notz}}]{SantnerEtAl2003}
\bibinfo{author}{\bibfnamefont{T.~J.} \bibnamefont{Santner}},
  \bibinfo{author}{\bibfnamefont{B.~J.} \bibnamefont{Williams}},
  \bibnamefont{and} \bibinfo{author}{\bibfnamefont{W.~I.} \bibnamefont{Notz}},
  \emph{\bibinfo{title}{{T}he design and analysis of computer experiments}}
  (\bibinfo{publisher}{Springer, New York}, \bibinfo{year}{2003}).

\bibitem[{\citenamefont{Clerouin et~al.}(2016)\citenamefont{Clerouin, Desbiens,
  and Arnault}}]{ClerouinEtAl2016a}
\bibinfo{author}{\bibfnamefont{J.}~\bibnamefont{Clerouin}},
  \bibinfo{author}{\bibfnamefont{N.}~\bibnamefont{Desbiens}}, \bibnamefont{and}
  \bibinfo{author}{\bibfnamefont{P.}~\bibnamefont{Arnault}},
  \bibinfo{journal}{Physical Review Letters submitted}  (\bibinfo{year}{2016}).

\end{thebibliography}
\bibliographystyle{apsrev.bst}


\end{document}